\documentclass[structabstract]{aa}  
\usepackage{graphicx}
\usepackage{txfonts}
\usepackage{longtable}

\begin{document}
   \title{EVN observations of 6.7~GHz methanol masers in clusters of massive young stellar objects.
   \thanks{Figures~\ref{distrib2}, \ref{gauss}, \ref{sed-on} and Tables~\ref{table3} and \ref{sedin} are 
only available in electronic form via http://www.edpsciences.org} 
}

   \author{A. Bartkiewicz
           \inst{1}, 
           M. Szymczak
           \inst{1},
           \and
	   H.J. van Langevelde
           \inst{2,3}}

       \institute{Centre for Astronomy, Faculty of Physics, Astronomy and
         Informatics, Nicolaus Copernicus University, Grudziadzka 5, 87-100 Torun, Poland, 
\email{[annan;msz]@astro.uni.torun.pl}
\and      Joint Institute for VLBI in Europe, Postbus 2, 7990 AA
          Dwingeloo, The Netherlands, \email{langevelde@jive.nl}
\and      Sterrewacht Leiden, Leiden University, Postbus 9513, 2300 RA Leiden, The Netherlands
             }

   \date{Received 9 September 2013; accepted 31 January 2014}

 
  \abstract
{Methanol masers at 6.7~GHz are associated with high-mass star-forming
regions (HMSFRs) and often have mid-infrared (MIR) counterparts characterized by 
extended emission at 4.5$\mu$m, which likely traces outflows from massive 
young stellar objects (MYSOs).}
{Our objectives are to determine the milliarcsecond (mas) morphology of the maser emission
and to examine if it comes from one or
several candidate MIR counterparts in the clusters of MYSOs.}
{The European VLBI Network (EVN) was used to image the 6.7\,GHz maser line with $\sim$2\farcm1 field
of view toward 14 maser sites from the Torun catalog. 
Quasi-simultaneous observations were carried out with the Torun 32\,m telescope.}
{We obtained maps with mas angular resolution that showed diversity
of methanol emission morphology: a linear distribution (e.g.,
G37.753$-$00.189), a ring-like (G40.425$+$00.700), and a complex
one (e.g., G45.467$+$00.053). 
The maser emission is usually associated with the strongest MIR counterpart in
the clusters; no maser emission was detected from other MIR sources in the fields of view of 2\farcm1
in diameter.  
The maser source luminosity seems to correlate with the total luminosity of the central MYSO.
 Although the Very Long Baseline Interferometry (VLBI) technique resolves a significant part of the maser
emission, the morphology is still well determined. This indicates that
the majority of maser components have compact cores.}
{}

\titlerunning{Milliarcsecond structure of methanol masers}
 
\authorrunning{A. Bartkiewicz, M. Szymczak \and H.J. van Langevelde}

   \maketitle

\section{Introduction}
The 6.7\,GHz methanol maser emission is widely assumed to be associated with massive young stellar objects
(MYSOs), but 
it is still unclear which structures it probes in the circumstellar environment. High angular resolution studies have 
revealed quite diverse morphologies of methanol maser sources {from simple 
and linear to curved and complex, or even circularly symmetric ones (e.g., 
Minier et al.~\cite{minier00}; Dodson et al.~\cite{dodson04}; Bartkiewicz et
al.~\cite{bartkiewicz09}; Pandian et al.~\cite{pandian11}; Fujisawa et
al.~\cite{fujisawa14}). It has been suggested that linear maser structures 
with velocity gradients indicate circumstellar disks seen
edge-on (Norris et al.\,\cite{norris98}; Minier et al.~\cite{minier00}), and 
the velocity gradients within individual maser clouds perpendicular to the major 
axis of maser distribution point planar shocks (Dodson et
al.\,\cite{dodson04}). 
Arched and ring-like structures can be explained by models
of rotating and expanding disks or outflows where the maser arises at the interface between disk/torus 
and a flow (Bartkiewicz et al.~\cite{bartkiewicz09}; Torstensson et al.~\cite{torstensson11}).
Detailed proper motion studies of methanol emission, 
done in only a few sources so far, revealed different scenarios of ongoing phenomena.  
Observations of the 3D velocity field of methanol masers in the protostellar AFLG~5142 proved that 
the emission arises in the infalling gas of a molecular envelope with a 300\,AU radius (Goddi et al.\,\cite{goddi11}).
Sanna et al.~(\cite{s10a}) noticed rotation of methanol masers around a central mass
in G16.59$-$0.05. Towards G23.01$-$0.41 a composition of slow radial
expansion and rotation motions were detected (Sanna et al.~\cite{s10b}). In
IRAS~20126$+$4104 methanol maser spots are associated with the circumstellar disk 
around the object, and also trace the disk at the interface with the bipolar jet
(Moscadelli et al.~\cite{m11a}). 

Cyganowski et al.~(\cite{cyganowski09}) found that 6.7~GHz methanol masers frequently appear
toward mid-infrared (MIR) sources with extended emission at 4.5$\mu$m,
 the so-called 
{\it extended green objects} (EGOs, Cyganowski et al.\,\cite{cyganowski08}) 
or {\it green fuzzies} (Chambers et al.~\cite{chambers08}). 
This emission probably signposts shocked molecular gas, 
mainly H$_2$ and CO molecules, related to protostellar outflows. 
However, detailed studies of each target may also identify its "falsely
appearing {\it green}" (De Buizer \& Vacca~\cite{debuizer10})}. 
Simple analysis of the {\it Spitzer} 
GLIMPSE data (Benjamin et al.\,\cite{benjamin03}; Fazio et
al.\,\cite{fazio04})   
maps proved that the majority of methanol masers 
from Bartkiewicz et al.~(\cite{bartkiewicz09}) are spatially associated within 1\arcsec\, with EGOs and there is 
a trend that the regular maser structures coincide with the strongest MIR objects. 

In this paper we report on the European VLBI Network (EVN) observations of a sample of 6.7~GHz maser sources. 
Our aims are threefold: i) to obtain
the accurate absolute position of the sources and to determine their morphology, ii) to check if the emission 
is associated with a single or with several MYSOs, and iii) to determine if the
maser morphology is consistent with an outflow.
With the new data we discuss the range of physical parameters of central objects which power the methanol maser emission.  

\section{Observations and data reduction}
\subsection{Sample selection}
The maser targets for the VLBI observations were selected from the Torun
32\,m telescope archive data of methanol maser lines (the catalog was published
only recently by Szymczak et al.~\cite{szymczak12}),
on the basis of the significant velocity extent ($>$8\,km\,s$^{-1}$) of the 
6.7\,GHz emission with multiple features of flux density greater than 2\,Jy. 

We assumed that these characteristics are typical for sources with
complex spatial maser morphology.
In addition, preference was given to objects with mid-infrared emission
(MIR) counterparts in clusters of size
2$-$3\arcmin\ characterized by extended emission at 4.5$\mu$m.
The MIR emission morphology was examined using the GLIMPSE images retrieved
from the Spitzer
archive\footnote{http://irsa.ipac.caltech.edu}. 
Images of area 5\farcm5$\times$5\farcm5 centered at the location of maser
sources (Szymczak et al.\,\cite{szymczak12}) 
were loaded into the Astronomical Image Processing 
System (AIPS) developed by the National Radio Astronomy Observatory (NRAO). 
Maps of the 4.5$\mu$m$-$3.6$\mu$m emission excess
were created by subtracting the 3.6$\mu$m image
from the 4.5$\mu$m image of each maser site. The candidate site was selected if
it contains at least one object with
the 4.5$\mu$m$-$3.6$\mu$m emission excess extended $\ge$15\arcsec at a
surface brightness $\ge$2\,MJy\,sr$^{-1}$.
We note, this procedure does not meet the more stringent criteria used to
search for EGOs (Cyganowski et al.\,\cite{cyganowski09}).

The list of targets included ten star forming sites
selected from the Torun methanol maser catalog 
(Table~\ref{table1}). Inspection of data obtained 
with the Arecibo dish (Pandian et al.\,\cite{pandian07b}) revealed that
G41.12$-$0.22 and G41.16$-$0.18 were two different targets (not resolved
with the 5\farcm5 beam of the 32\,m telescope), while G37.76$-$0.21 and
G43.16$+$0.01, contained multiple maser sources. Therefore,  
the EVN observations were done for 14 pointing positions (Table~\ref{table1}).
The pointing was done toward bright MIR counterparts.

\begin{table*}
\caption{Details of EVN observations.}
\label{table1}       
\centering            
\begin{tabular}{lllcccc}
\hline\hline
Source$^{\star}$ & \multicolumn{2}{c}{Pointing positions (J2000)} & Phase-calibrator & Separation
& Observing & Synthesized beam\\
Gll.ll$\pm$bb.bb& RA (h m s) & Dec ($^{\rm o}$ ' '') &  & (\fdg) & run$^{\star\star}$ &
(mas$\times$mas,$^{\rm o}$)\\
\hline
G37.76$-$00.21$^{2}$  & 19 00 55.4 & 04 12 12.5 & J1907$+$0127 & 3.17 & 2 & 5.4$\times$5.1;$-$30\\
               & 19 01 02.0 & 04 12 01.7 & J1907$+$0127 & 3.15 & 2 & --\\

G40.28$-$00.22$^{1}$ & 19 05 41.2 & 06 26 12.5 & J1912$+$0518 & 2.12 & 2 & 5.3$\times$5.1;$+$59\\
G40.43$+$00.70$^{1}$  & 19 02 39.6 & 06 59 09.1 & J1912$+$0518 & 3.05 & 1 & 5.6$\times$4.7;$-$49\\
G41.12$-$00.22 & 19 07 14.8 & 07 11 00.7 & J1912$+$0518 & 2.35 & 2 & 5.3$\times$5.0;$+$54\\
G41.16$-$00.18 & 19 07 11.2 & 07 14 04.4 & J1912$+$0518 & 2.40 & 2 & 5.3$\times$4.7;$+$55\\
G41.23$-$00.20$^{2}$  & 19 07 21.4 & 07 17 08.4 & J1912$+$0518 & 2.40 & 2 & 5.4$\times$4.9;$+$54\\ 
G41.35$-$00.13$^{1}$  & 19 07 21.7 & 07 25 17.7 & J1912$+$0518 & 2.53 & 1 & 5.2$\times$4.9;$-$78\\
G43.16$+$00.01$^{1}$  & 19 10 11.4 & 09 07 06.2 & J1912$+$0518 & 3.88 & 1 & --\\
(W49N)                 & 19 10 14.0 & 09 05 53.3 & J1912$+$0518 & 3.86 & 1 & --\\
                       & 19 10 14.7 & 09 06 16.8 & J1912$+$0518 & 3.86 & 1 & 6.1$\times$5.1;$-$53\\
G45.47$+$00.05$^{1}$  & 19 14 25.6 & 11 09 27.0 & J1925$+$1227 & 3.05 & 2 & 5.3$\times$4.7;$+$64\\
G45.47$+$00.13$^{1}$  & 19 14 07.2 & 11 12 15.4 & J1925$+$1227 & 3.10 & 2 & 5.3$\times$4.7;$+$65\\
G59.78$+$00.07$^{1}$  & 19 43 11.1 & 23 44 03.0 & J1931$+$2243 & 2.89 & 1 & 6.5$\times$3.8;$-$55\\
\hline
\end{tabular}
\tablefoot{$^{\star}$ Names are the Galactic coordinates derived from the Torun survey 
           (Szymczak et al.\,\cite{szymczak12}). 
           $^{\star\star}$ 2010 March 14 (run 1), 2010 March 15 (run 2).
           The intervals between the EVN session and the single-dish observations were 
           $^{(1)}$1-4 weeks and $^{(2)}$14-15 months.
} 
\end{table*}  

\subsection{EVN observations}
The EVN\footnote{The European VLBI Network is a joint facility of European, Chinese, South
African, 
and other radio astronomy institutes funded by their national research councils.} 
observations of 10 regions, using 14 tracking centers
(Table~\ref{table1}) were carried out at 6668.519~MHz on 2010 March 14 (run 1) and 15 (run 2) for 10~hr each 
(program code: EB043). The following antennas were used: 
Jodrell Bank, Effelsberg, Medicina, Onsala, Noto, Torun, Westerbork, and 
Yebes. The Torun antenna was not used in run 2.
A phase-referencing scheme was applied with reference sources as listed in Table~\ref{table1}
in order to determine the absolute positions of the targets at the
level of a few mas (Bartkiewicz et al.~\cite{bartkiewicz09}). We 
used a cycle time between the maser and phase-calibrator of 195~s$+$105~s. This yielded
a total integration time for each individual source of $\sim$50~min and $\sim$40~min in runs 1 and 2, 
respectively. The bandwidth was set to 2\,MHz yielding 90\,km\,s$^{-1}$ velocity coverage 
(covering the local standard of rest (LSR) velocity range from
18~km~s$^{-1}$ to 108~km~s$^{-1}$ or from $-$30~km~s$^{-1}$ to
60~km~s$^{-1}$ as listed in Table~\ref{table_EGO}).
Data were correlated with the Mk\,IV Data Processor operated by JIVE with 1024 spectral channels. 
The resulting spectral resolution was 0.089\,km\,s$^{-1}$. 
The data integration was kept to 0.25~s, to minimize time smearing
in the {\it uv}-plane. One can estimate
that in this way we can use a $\sim$2\farcm1 field of view, over which the
response to point sources is degraded
by less than $\sim$10\%.

The data calibration and reduction were carried out with  AIPS using standard procedures for spectral line observations. We used the Effelsberg antenna 
as a reference. The amplitude was calibrated through measurements of the system temperature at each 
telescope and application of the antenna gain curves. The parallactic angle corrections were subsequently added 
to the data. The source 3C345 was used as a delay, rate, and bandpass calibrator. The phase-calibrators 
J1907$+$0127, J1912$+$0518, J1925$+$1227, and J1931$+$2243 were imaged and flux densities of 170, 
163/132 (run 1/2), 112, and 259\,mJy were obtained, respectively. 
The maser data were corrected for the effects of the Earth's rotation and
its motions within the solar system and toward the LSR. 

In order to find the positions of the emission for each target (as we did not obtain consistent results from the
fringe-rate mapping, probably because of the closeness to 0\degr \,\,declination), we inspected the vector-averaged spectra 
by shifting the phase-centers from $-$2\arcmin\, to $+$2\arcmin\, in right ascension and declination by  
1\arcsec\, steps. A verification criterion
for a given spectrum was based on maximizing the intensity for a given maser feature 
and verifying that its phase was close to 0\degr. When the maximum emission feature was identified we created a dirty map
of size 8\arcsec$\times$8\arcsec\, at the given position and the more
accurate coordinates of emission were estimated with a similar production of a smaller
image (1\arcsec$\times$1\arcsec). We then we ran a 
self-calibration procedure with a few cycles shortening the time interval to
4~min using the clean components 
of the compact and bright maser spot map. The first, dirty map was applied as a model in order to avoid shifting the position of a dominant component and losing its absolute position. 
Finally, naturally-weighted maps of spectral channels were produced over the velocity range where 
the emission was seen in the scalar-averaged spectrum. A pixel separation of 1\,mas in both coordinates was
used for the imaging.  The resulting synthesized beams are listed in Table~\ref{table1}. The rms noise level 
(1$\sigma_{\rm rms}$) in emission line-free channels was typically from
6~mJy to 22~mJy for each source (Table~\ref{table_EGO}). In the case when
not all maser features seen in the scalar-averaged spectrum were recovered from the maps using task ISPEC, 
we restarted a search for emission at the given velocity in a way similar to the above. 

The positions of the methanol maser spots in all channel maps were determined by fitting two-dimensional Gaussian 
models. The formal fitting errors resulting from the beamsize/signal-to-noise ratio were less than 0.1\,mas. 
The absolute position accuracy of maser spots was estimated in Bartkiewicz et al. (\cite{bartkiewicz09})
to be a few mas. 

In two targets, G41.16$-$00.20 (with the intensity of the brightest spot of 0.65~Jy~beam$^{-1}$) and 
G45.493$+$00.126 (4.15~Jy~beam$^{-1}$), 
the phase-referencing failed and the positions of these sources are less accurate.
The offsets of the emission from the tracking centers were significant, (46\farcs9,
$-$46\farcs4) and (61\farcs2, 51\arcsec) in right ascension and 
declination, respectively. This made the images too noisy for reliable
identification of spots without an additional phase-calibration with FRING. 
Inspection of the dirty 
maps of the first source implied $\pm$1" and $\pm$3" measurement uncertainties in right ascension and declination, respectively, 
whereas for the second source the uncertainties were $\pm$0\farcs1 in each coordinate.
These are maximum values because our comparison with the EVLA and MERLIN data (Pandian et al.\,\cite{pandian11}) imply
the positional differences of 0\farcs30 and 0\farcs06 only for G41.16$-$00.20 and G45.493$+$00.126,
respectively.

\subsection{Single-dish observations}
Spectra at 6.7\,GHz of nine sites were taken with the Torun 32m telescope as a part of a monitoring program 
described in Szymczak et al.\,(\cite{szymczak12}).
For most objects the spectra were obtained at offsets of 1$-$4 weeks from the EVN session (Table~\ref{table1}).
The spectral resolution was 0.04\,km\,s$^{-1}$, the typical sensitivity was 0.6\,Jy, and 
the flux density calibration accuracy was about 15\%.   

\section{Results}
A total of 15 maser sources were successfully mapped. Table~\ref{table2} lists the updated names based
on the newly determined positions, the coordinates, the velocity ($V_{\rm
p}$), and the
intensity ($S_{\rm p}$) of the brightest spot of each source, the velocity extent of emission ($\Delta
V$), as well as the maser extent along
the major axis (the longest distance between two maser spots within a given
target) and the number of measured spots.

\begin{table*}
\caption{Results of EVN observations.}
\label{table2}       
\centering            
\begin{tabular}{lllcccccc}
\hline\hline
Source & \multicolumn{2}{c}{Position of the brightest spot (J2000)} & V$_{\rm p}$ & $\Delta$V & S$_{\rm p}$ & \multicolumn{2}{c}{Number of } & Extent\\
Gll.lll$\pm$bb.bbb& RA (h m s) & Dec ($^{\rm o}$ ' '') & (km s$^{-1}$) & (km s$^{-1}$) & (Jy\,beam$^{-1}$)& spots &
clouds$^{(3)}$ & (arcsec (AU)$^{(4)}$)\\
\hline
G37.753$-$00.189 & 19 00 55.421 & 04 12 12.5405 & 55.0 & 10.8 & 0.33 & 16 & 4 & 0.16 (1410)\\
G40.282$-$00.219 & 19 05 41.215 & 06 26 12.7034 & 74.4 & 18.4 & 7.20 & 123 & 15 & 0.55 (2710) \\ 
G40.425$+$00.700 & 19 02 39.620 & 06 59 09.0686 & 16.0 & 11.0 & 12.74 & 127 & 17 & 0.36 (4100) \\
G41.123$-$00.220 & 19 07 14.856 & 07 11 00.6593 & 63.4 & 8.9 & 2.03 & 19 & 3 & 0.05 (435)\\ 
G41.16$-$00.20$^{(1)}$ & 19 07 14.35  & 07 13 18.0 & 61.8 & 7.0 & 0.65 & 11 & 2 & 0.055 (480) \\
G41.226$-$00.197 & 19 07 21.378 &  07 17 08.1392 & 57.0 & 8.0 & 1.50 & 65 & 8 & 0.06 (520) \\
G41.348$-$00.136 & 19 07 21.839 & 07 25 17.6318 & 12.3 & 7.9 & 7.31 & 66 & 9 & 0.075 (870)\\  
G43.165$+$00.013 (W49N) & 19 10 12.882 & 09 06 12.2299 & 9.3 & 11.8 & 6.19 & 83 & 12 & 0.18 (2000) \\ 
G43.171$+$00.004 (W49N) & 19 10 15.353 & 09 06 15.4321 & 19.0 & 3.4 & 1.30 & 36 & 5 & 0.20 (2222)\\ 
G43.167$-$00.004 (W49N) & 19 10 16.720 & 09 05 51.2556 & -1.2 & 0.3 & 0.23 & 4 & 0 & 0.004 (44)\\
G43.149$+$00.013 (W49N) & 19 10 11.048 & 09 05 20.5179 & 13.2 & 1.0 & 1.45 & 10 & 3 & 0.19 (2111) \\
G45.467$+$00.053 & 19 14 24.147 & 11 09 43.4140 & 56.0 & 3.9 & 3.02 & 52 & 8 & 0.03 (230)\\ 
G45.473$+$00.134 & 19 14 07.361 & 11 12 15.9570 & 65.8 & 6.9 & 3.67 & 39 & 6 & 1.06 (7310) \\ 
G45.493$+$00.126$^{(2)}$ & 19 14 11.356 & 11 13 06.353 & 57.2 & 1.2 & 4.15 & 14 & 2 & 0.003 (21)\\ 
G59.782$+$00.065 & 19 43 11.247 & 23 44 03.2870 & 27.0 & 13.3 & 37.1 & 170 & 24 & 1.25(2750)\\
\hline
\end{tabular}
\tablefoot{$^{(1)}$Position from the VLA data: RA(J2000) = $19^{\rm h}07^{\rm m}14\fs369$, Dec(J2000) = 07\degr13\arcmin18\farcs08 and
           $^{(2)}$position from the MERLIN data: RA(J2000) = $19^{\rm h}14^{\rm m}11\fs357$, Dec(J2000) = 11\degr13\arcmin06\farcs41
           (Pandian et al.\,\cite{pandian11}).
           $^{(3)}$Clouds with Gaussian velocity profiles only. $^{(4)}$
           Linear maser extent calculated for the distances as listed in
           Table~\ref{source-prop}.
} 
\end{table*}  

\begin{figure*}
\centering
    \includegraphics[scale=0.75]{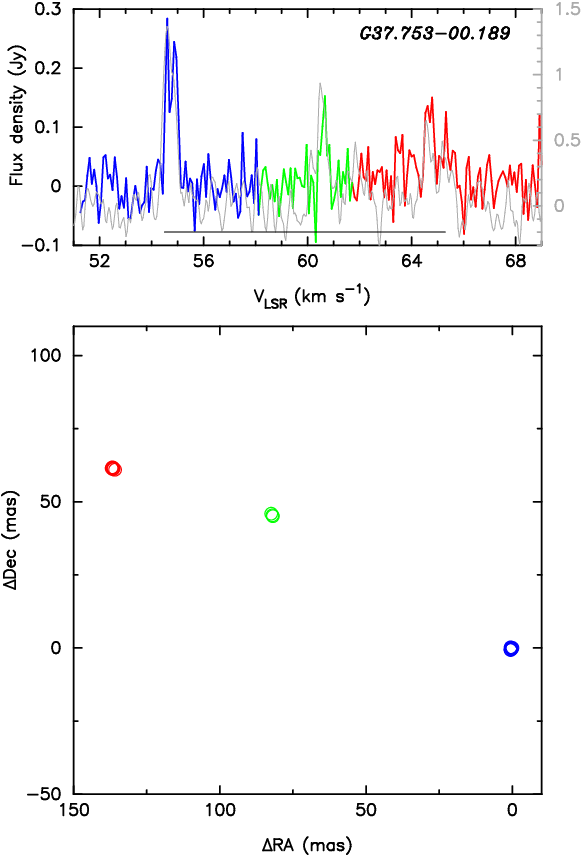}
    \includegraphics[scale=0.75]{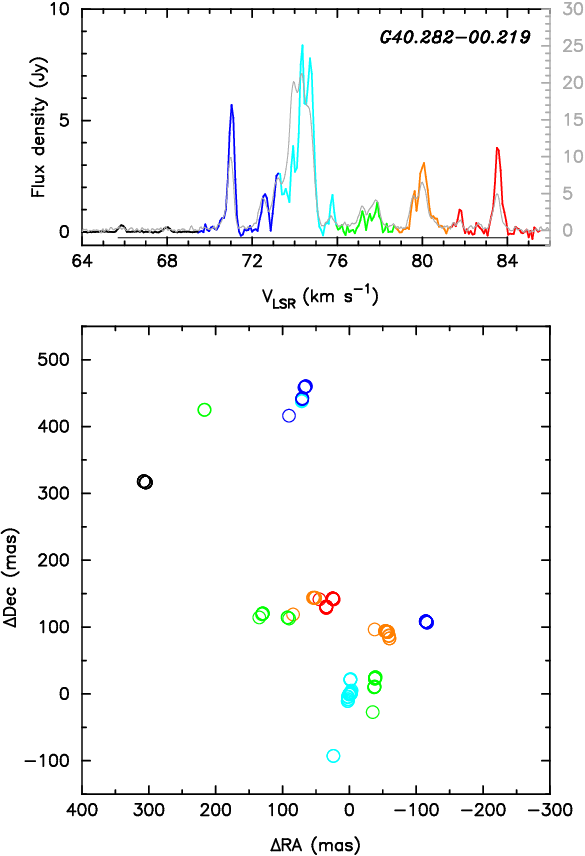}
    \includegraphics[scale=0.75]{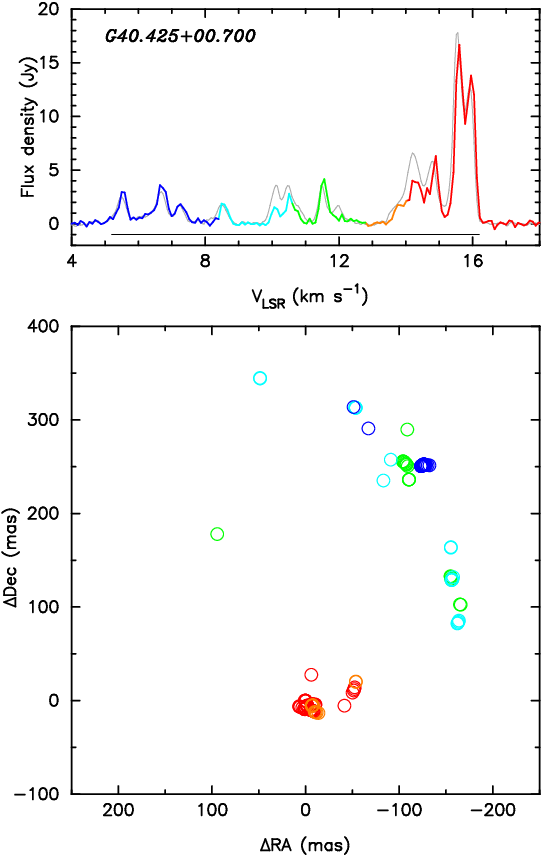}
    \includegraphics[scale=0.75]{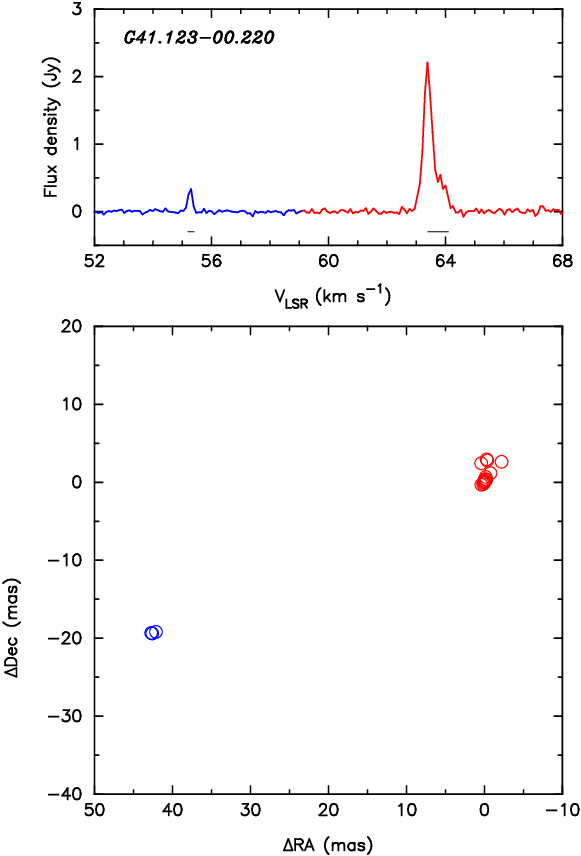}
    \caption{Spectra and maps of 6.7~GHz methanol masers detected using the EVN.
    The names are the Galactic coordinates of the brightest spots listed
    in Table~\ref{table2}. The colors of circles relate to the LSR velocities as shown in the spectra. 
    The map origins are the locations of the brightest spots
    (Table~\ref{table2}). The gray lines show the Torun 32~m dish spectra.
    If needed, the separate scale of the flux density is presented on the
    left (EVN) and right (Torun) sides. The thin bars under the spectra
    show the LSR velocity ranges of spots displayed. The plots for the
    remaining targets are presented in the on-line material (Appendix~A).} 
    \label{distrib}
    \end{figure*}

In Figures~\ref{distrib} and \ref{distrib2} we present the distribution of the maser emission and the spectrum for each source.
The spectra were extracted using the AIPS task ISPEC from the image datacubes using the smallest region 
covering the entire emission. For two maser sites, W49N and
G59.783$+$00.065, where the emission comes from
maser groups separated by more than 0\farcs5, but overlaps in velocity, the presented spectra are the sum of
the fluxes from all regions. The single-dish spectra are displayed when available.

A maser {\it cloud} was defined when the emission in at least three consecutive spectral channels  
coincides in position within half of the synthesized beam ($\sim$2.5\,mas).
We summarize the numbers and details of clouds with Gaussian velocity
profiles for each target in Tables~\ref{table2} and \ref{table3}, respectively.
 The velocity ($V_{\rm fit}$), line width at half 
maximum (FWHM), and flux density ($S_{\rm fit}$) were obtained by fitting a Gaussian profile to the spectrum.
The projected length ($L_{\rm proj}$) of maser clouds and, if seen, the velocity gradient ($V_{\rm
grad}$) are given. The Gaussian velocity profiles are plotted in Fig.~\ref{gauss}
({\it on-line}). 

All the detected maser sources listed in Table\,\ref{table2} coincide within
$<$0\farcs5 with MIR counterparts,
which are listed in Table\,\ref{table_EGO}.
As the field of view of EVN observations was about 2\farcm1\, we searched
for the maser emission toward other
MIR objects, visible in the GLIMPSE maps (Benjamin et
al.\,\cite{benjamin03}), lying within a radius of 60\arcsec\,
of the phase centers (Table~\ref{table1}). No new emission at 6.7\,GHz was
found within the observed LSR velocity 
range using the searching method described in Sect.~2.2.
The names, Galactic coordinates, 1$\sigma$ noise levels, velocity range
searched for the maser emission,
IRAC colors [3.6]$-$[4.5] and [4.5]$-$[5.8], and luminosities of MIR
counterparts of maser and non-maser objects
are given in Table\,\ref{table_EGO}. The positions and MIR flux densities
are taken from GLIMPSE I Spring '07 Archive via Gator in the NASA/IPAC Infrared
Science Archive. 
The crowded region of W49N is omitted in Table\,\ref{table_EGO} because of
difficulty of identification
of discrete sources; all MIR counterparts of the four maser sources are
blended and/or saturated. 
In Table\,\ref{table_EGO} only two MIR objects G40.2819-0.2197 and
G45.4725$+$0.1335 classified as ``possible''
MYSO outflow candidates in the EGO catalog (Cyganowski et
al.\,\cite{cyganowski08}) are associated with the maser
emission. One object G45.4661$+$0.0457 listed in Cyganowski et al. as
``likely'' MYSO outflow candidate has no 6.7\,GHz maser emission.

\subsection{Individual sources}
{\bf G37.753$-$00.189.} The weakest source in the sample with a brightest spot of 334~mJy at 
a velocity of 55.0~km~s$^{-1}$. In total we registered 16 spots, over a  velocity range from 54.5 to
65.3~km~s$^{-1}$, grouped into three clusters distributed in a linear structure of length  
$\sim$150\,mas with a clear velocity gradient, redshifted spots in the
northeast and blueshifted ones in the southwest. The blue-shifted (54.5$-$55.0\,km\,s$^{-1}$) cluster  
contains two clouds blended in velocity and coinciding in position within $\sim$0.6\,mas. 
Pandian et al.~(\cite{pandian11}) failed to find any emission using MERLIN in
2007 (they did not notice any emission in the cross-spectra). 
However, they were successful when using the EVLA in A configuration in 2008; the emission from a compact group 
of spots with a peak of 2.1\,Jy in the velocity range of 54.5$-$55.1\,km\,s$^{-1}$ coincides within 
20\,mas with the EVN position, whereas weaker emission at velocities higher than 60.3\,km\,s$^{-1}$ 
is distributed over an area of 150$\times$80\,mas. This diffuse emission is seen with the EVN 
as two compact clouds of flux density lower than 0.15\,Jy. The integrated flux density obtained with
the VLBI is a factor of 4.9 lower than that observed with the 32\,m telescope.

{\bf G40.282$-$00.219.} This source with 123 maser spots detected between
65.6~km~s$^{-1}$ and 84~km~s$^{-1}$, 
shows a complex morphology. Fifteen clouds with Gaussian velocity profiles are distributed over an area
of 0\farcs4$\times$0\farcs6. We do not notice any overall regularity in
the LSR velocities, but 14 out of 15 identified maser clouds showed
individual velocity gradients
(Table~\ref{table3}). The source morphology seen with the EVLA by Pandian et
al.~(\cite{pandian11}) 
is very similar to what we find here, with the exception that they did not map any emission at velocities 
higher than 79~km~s$^{-1}$ because of the limit of the bandwidth. The shape of the spectrum from the EVN
generally agrees with that from the 32~m dish, but the flux density of individual features appears 
to be reduced by a factor of 1.3$-$6.7. The features at 65.8~km~s$^{-1}$ and 68.0\,km\,s$^{-1}$ detected only in
the VLBI observation were at the 3$\sigma_{\rm rms}$ noise level in the
single-dish spectrum.

{\bf G40.425$+$00.700.} The 127 spots detected in the velocity range of
5.2~km~s$^{-1}$ to 16.2~km~s$^{-1}$ delineate
an arched structure of a size of 350\,mas; 111 of them formed 16 clouds with
Gaussian velocity profiles. The strongest and most redshifted ($>$13.5\,km\,s$^{-1}$) emission
forms the southern cluster of eight clouds distributed over an area of 60\,mas. 
The extreme blueshifted ($<$8\,km\,s$^{-1}$) emission is clustered at the northern side of
the arch.
The source morphology resembles that reported for the ring source G23.657$-$00.127 (Bartkiewicz et al.~\cite{bartkiewicz05}). 
The spectrum obtained with the EVN is very similar to the single-dish spectrum; for two features only
the emission is resolved out by $\sim$30$-$40\%. Caswell et al.~(\cite{caswell09}) reported a velocity 
range and peak flux similar to ours, but their coordinates differ by 1\farcs5 in
declination. However, the new position agrees with that obtained by the BeSSeL 
project\footnote{www.bessel2.vlbi-astrometry.org} (Brunthaler et
al.~\cite{brunthaler11}).

{\bf G41.123$-$00.220.} Two maser clusters are separated by 50\,mas and differ in velocity 
by 8.9\,km\,s$^{-1}$. The emission near 55.3\,km\,s$^{-1}$ comes from one eastern cloud, whereas that near 
63.4$-$63.9\,km\,s$^{-1}$ comes from the western cluster of size $\sim$5\,mas. 
Our map is generally consistent with that obtained with the EVLA (Pandian et al.~\cite{pandian11}), but the western
cluster appears to contain diffuse emission.

{\bf G41.16$-$00.20.} Weak emission was found significantly shifted 
from the phase center so fringe fitting was applied (Sect.\,2.2). Two maser clouds are separated 
by 50\,mas and their velocities differ by 6~km~s$^{-1}$. The source observed with the EVLA shows a complex structure
($\sim$100$\times$50\,mas in extent) of the eastern and redshifted (61.6$-$63.7\,km\,s$^{-1}$) emission
(Pandian et al.~\cite{pandian11}).
The flux density recovered with the EVN was about a factor of 2-3 lower than that detected with the EVLA.
 With this characteristic we can conclude that the VLBI has resolved
most of its emission.

{\bf G41.226$-$00.197.} Sixty-five spots were detected above 74~mJy in the velocity range of 55.0$-$63.0~km~s$^{-1}$. 
They form eight clouds with Gaussian velocity profiles in four clusters. The overall morphology is very similar 
to that observed with MERLIN (Pandian et al.~\cite{pandian11}), with the exception of the two southern redshifted 
clusters which are more compact. The single-dish spectrum has a similar shape
to the MERLIN spectrum, but it is
clear that the emission is resolved out by a factor of 2$-$3 in the EVN observation.
 The redshifted emission is located in the south, while the blueshifted
in the north.

{\bf G41.348$-$00.136.} The 66 spots detected in the velocity range from
6.8~km~s$^{-1}$ to 14.7~km~s$^{-1}$ forms nine clouds with
Gaussian velocity profiles and are clustered into two groups. The southeastern cluster is blueshifted ($<$10\,km\,s$^{-1}$), 
while the northwestern cluster is redshifted. The two clusters are separated by $\sim$50\,mas. 
The overall morphology of emission agrees well with that obtained with MERLIN (Pandian et al.\,\cite{pandian11}). 
For most components the flux density is slightly (0.5$-$1.0\,Jy) reduced compared to the
single-dish and
MERLIN spectra, but the component near 12~km~s$^{-1}$ is lowered by a factor of two.

{\bf W49N.} This maser site is of special interest since four sources (Table\,\ref{table2}) were detected over 
an area of 84\arcsec$\times$55\arcsec in the velocity range from
$-$1.2~km~s$^{-1}$ to 22.2~km~s$^{-1}$. Two southern
sources, 
G43.149$+$00.013 and G43.167$-$00.004, show linear structures (PA = $-$45\degr) composed of only ten and four spots, 
respectively with respective lengths of 4.5\,mas and 183\,mas. Both sources exhibit a clear velocity gradient.
The morphology of G43.149$+$00.013 agrees well with that observed with MERLIN (Pandian et al.~\cite{pandian11}) 
while the weakest source G43.167$-$00.004 is largely resolved with the EVN. We note very narrow emission of 0.3
(in the LSR velocity range from $-$1.25~km~s$^{-1}$ to $-$0.99\,km\,s$^{-1}$) and 1.1\,km\,s$^{-1}$ (from
13.06~km~s$^{-1}$ to 14.11\,km\,s$^{-1}$) in both sources.
The other two sources G43.165$+$00.013 and G43.171$+$00.004 are complex with 83 and 36 spots spread over areas of
120~mas $\times$ 190~mas and 120~mas $\times$ 180~mas, respectively. The velocity extent of emission from G43.165$+$00.013
is 11.9\,km\,s$^{-1}$, while that from G43.171$+$00.004 is 3.4\,km\,s$^{-1}$. The overall morphologies of these
four objects agree with those obtained with MERLIN (Pandian et al.~\cite{pandian11}).
We did not find any emission from source G43.18$-$0.01 that was detected in the 
Arecibo survey (Pandian et al.\,\cite{pandian07b}) and imaged with one
spot only (with a flux density of 0.32~mJy) using MERLIN (Pandian et
al.\,\cite{pandian11}).
The EVN observation poorly recovered the maser emission from the site; the flux density of individual components 
are a factor of 3-7 lower that observed with the single dish.

{\bf G45.467$+$00.053.} Emission of 4\,km\,s$^{-1}$ width, composed of 52 spots clustered in
eight groups, was detected
over an area of 20~mas $\times$ 30~mas. The extreme blue- and redshifted emissions delineate two linear 
structures at PA=$-$45\degr with monotonic velocity gradients. The emission at the intermediate velocity of 
57.2$-$57.9~km~s$^{-1}$ forms a 15\,mas linear structure at PA=90\degr. Maps obtained with MERLIN show 
less complex morphology (Pandian et al.~\cite{pandian11}). 
The single-dish spectrum implies that some components at intermediate velocity were resolved out with EVN by a factor 
of two. No emission near 50~km~s$^{-1}$ was detected in the VLBI data.

{\bf G45.473$+$00.134.} Thirty-nine maser spots in the velocity range from
59.5~km~s$^{-1}$ to 66.5~km~s$^{-1}$ form three clusters separated 
by 0\farcs4$-$1\farcs1. The clusters with emission at velocity lower than  64~km~s$^{-1}$ lie 0\farcs7 
and 0\farcs4 to the southwest and to the north, respectively, from the strongest emission near 66\,km\,s$^{-1}$.
The overall morphology is similar to that obtained with MERLIN (Pandian et al.~\cite{pandian11}) and the flux density
of individual components are only slightly lower than the single-dish spectrum.

{\bf G45.493$+$00.126.} The 14 spots detected form a roughly linear structure of size 3\,mas with a monotonic velocity 
gradient of 0.38~km~s$^{-1}$~mas$^{-1}$ from the west to the east for the blue- and redshifted velocities, respectively.
A similar structure was observed with MERLIN (Pandian et al.~\cite{pandian11}). The
single-dish spectrum indicates that
only 50\% of the flux was recovered with the EVN for the emission at velocities higher than 57~km~s$^{-1}$.

{\bf G59.782$+$00.065.} Two elongated (290\,mas and 330\,mas) maser clusters separated by 860\,mas were detected. The eastern 
cluster is composed of ten clouds (71 spots) in the velocity range from
15.2~km~s$^{-1}$ to 27.6~km~s$^{-1}$ and the western cluster 
contains 14 clouds (99 spots) at velocities of 14.3$-$21.7~km~s$^{-1}$. Almost all maser spots (160/170) form
clouds which are fitted by Gaussian profiles. The single-dish spectrum indicates that 
for most of the components the flux density dropped by $\sim$30\% in the EVN observation. There are three features centered at 14.5,
21.5, and 24.5~km~s$^{-1}$ with flux densities which agree within less than 10\% of those measured with the
single-dish. 
The 12.2\,GHz methanol maser emission of only two features at
17.0~km~s$^{-1}$ and 26.9\,km\,s$^{-1}$ was reported to form two additional 
compact clusters of components separated by about 800\,mas (Minier et al.\,\cite{minier00}).

\section{Discussion}
In the subsequent analysis we will consider all the GLIMPSE sources that are listed in Table~\ref{table_EGO}. 
This allows us to consider sources with and without maser emission, although through the selection we are 
obviously biased toward maser sources. 

\subsection{Properties of maser clouds}
The kinematic distances to the targets are calculated with recipes of Reid et al. (\cite{r09}) assuming that 
the systemic velocity of each individual source is equal to the peak velocity of
the $^{13}$CO profile 
(Pandian et al.~\cite{pandian09}) or the middle velocity of the maser 
spectrum (Szymczak et al.\,\cite{szymczak12}). Distance ambiguities have been successfully resolved
toward  
all targets using the 21\,cm HI absorption line or the 6\,cm formaldehyde absorption line, taken from the
literature, and in a few cases the trigonometric distances have been adopted (Table\,\ref{source-prop}).

As we mentioned in Sect.~3, we have calculated the projected length of each
identified maser cloud as well as the velocity gradients
(Table\,\ref{table3}). Taking the
estimated distances we also list these values of the linear scales
(Table~\ref{table3}). In total, we identified 118 maser 
clouds with Gaussian velocity profiles. Seventeen features were fitted with at least 
two Gaussian profiles giving 29 additional profiles 
(e.g., clouds {\it 1}, {\it 2}, {\it 17} in G40.425$+$00.700; 
Table~\ref{table3} and Fig.~\ref{gauss} {\it on-line}).
The projected length of maser clouds ranges from 0.65\,AU to 113.32\,AU with a mean 
of 23.61$\pm$2.32\,AU and a median of 13.96\,AU. Velocity gradients of maser
clouds show values up to 0.59~km~s$^{-1}$~AU$^{-1}$, while in some cases 
no gradients were seen. The mean velocity gradient is
0.051$\pm$0.007~km~s$^{-1}$~AU$^{-1}$ and the median is 0.026~km~s$^{-1}$~AU$^{-1}$.
 The FWHM of all 147 (118$+$29) Gaussian profiles ranged from 
0.13~km~s$^{-1}$ to 1.3~km~s$^{-1}$ with a mean of 0.38$\pm$0.02\,km\,s$^{-1}$
and a median of 0.33~km~s$^{-1}$. These values are comparable to 
those reported from single-dish observations by 
Pandian \& Goldsmith (\cite{pandian07a}). We do not find 
a significant difference in median and average values of profile widths 
between nearby and distant objects. 

\begin{figure}
\centering
   \includegraphics{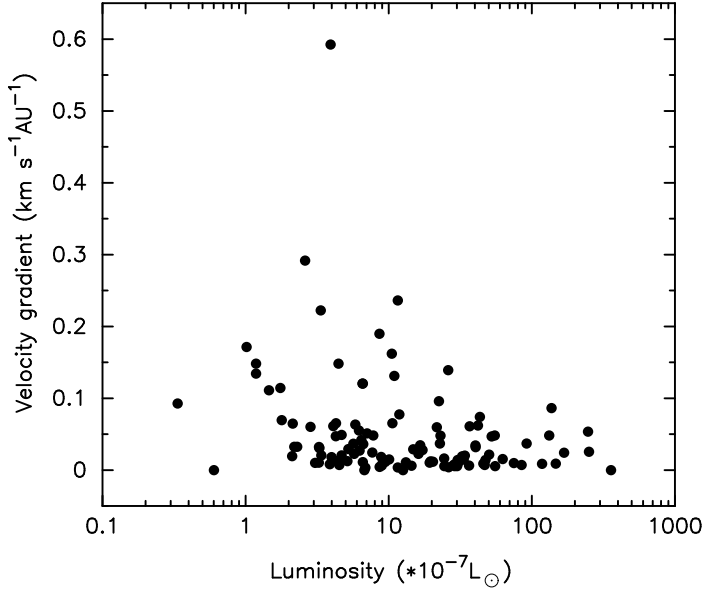}
    \caption{The relationship between the luminosity of a single methanol maser cloud and its velocity 
gradient (Table~\ref{table3}).}
    \label{gradient}
    \end{figure}

The luminosity of individual maser clouds was 
calculated according to the following formula:
$L_{6.7GHz}[L_{\sun}] = 6.9129 \times 10^{-9} D^{2}[\rm kpc] S_{int}[\rm
Jy~km~s^{-1}]$. Its range is $0.33-358\times 10^{-7} L_{\sun}$ and 
 has a mean of $28.5\pm4.1\times10^{-7} L_{\sun}$, and a median
of $10.79\times10^{-7} L_{\sun}$. As one can see the most of the clouds show
luminosity below $30 \times 10^{-7} L_{\sun}$ and a velocity gradient less
than 0.15~km~s$^{-1}$~AU$^{-1}$, but we also note that there is a tendency
for the high luminosity ($>$$30 \times 10^{-7} L_{\sun}$) clouds to have a gradient
velocity less than 0.1~km~s$^{-1}$~AU$^{-1}$ (Fig.~\ref{gradient}). 
This may be
related to the direction of the maser filament; the more it is aligned with the
line of sight, the less is the velocity gradient seen on the plane of the sky.

\subsection{Maser and YSO luminosities}
With our sample of maser sources we can attempt to estimate the physical parameters
of the central sources since the distances are quite reliably determined. 
The near- and mid-infrared fluxes for the counterparts of the 11 targets 
 (with the exception of W49N region because of 
its complexity in the VLBI maps and the lack of IR data
at sufficient resolution) were taken from the following publicly available data: 
UKIDSS-DR6 (Lucas et al.~\cite{lucas08}) or 2MASS All$-$Sky Point
Source Catalog (Skrutskie et al.~\cite{skrutskie06}) for G59.782$+$00.065, {\it Spitzer}  IRAC (Fazio et
al.~\cite{fazio04}) and MIPS (Rieke et al.~\cite{rieke04}), and MSX (Egan et al.~\cite{egan03}). 
The far-infrared and sub-millimeter fluxes toward  some of the targets
were found in the literature; they were taken with SCUBA (Di Francesco et
al.~\cite{difrancesco08}), LABOCA  and IRAM (Pandian et al.~\cite{pandian10}),
 BOLOCAM 
(searched for in Rosolowsky et al.~\cite{rosolowsky10} via the VizieR Service
and verified with the recent catalog by Ginsburg et
al.~\cite{gingsburg2013}), 
Herschel (Veneziani et
al.~\cite{veneziani13}), and SIMBA (Hill et al.~\cite{hill05}). We also made
 use of the RMS Database
Server\footnote{http://rms.leeds.ac.uk/cgi-bin/public/RMS$\_$ DATABASE.cgi}
(Urquhart et al.~\cite{urquhart08}) in order to verify the completeness
of found data for some targets. 
The source infrared fluxes are listed in Table \ref{sedin}.

\begin{figure}[t] 
\centering
    \includegraphics[scale=0.75]{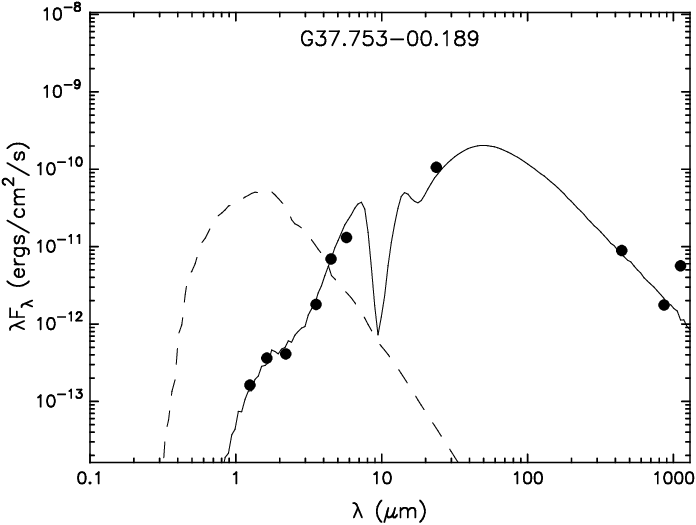}
    \caption{Fit to the spectral energy distribution of the first 6.7\,GHz source in the sample.
    The fits for the remaining targets are presented in the {\it on-line
    material}. The filled circles show the input
    fluxes (Table \ref{sedin}). The solid line shows the best fit and the gray lines show subsequent good fits. The dashed line 
    shows the stellar photosphere corresponding to the central source in the best fitting model in the absence of circumstellar
    extinction and in the presence of interstellar extinction.} 
    \label{fig-sed}
    \end{figure}

Assuming no variability of the infrared emission level, we applied the SED fitter developed by 
Robitaille et al. (\cite{robitaille07}) which is based on a grid of SED models spanning
a large range of evolutionary stages and stellar masses. They use young stellar objects with various combinations of 
circumstellar disks, infalling envelopes, and outflow cavities under the assumption that stars form via accretion through 
the disk and envelope. We set an uncertainty of distance parameter of either 10\% or that
listed in the literature in a source distance and varied the interstellar extinction, $A_V$, from 0 to 100
magnitudes. Figure \,\ref{fig-sed} shows the model fit of the SED for the first source in our
sample, G37.753$-$00.189. The SEDs of the remaining targets are presented on-line
(Fig.~\ref{sed-on}). For the sources 
G40.425$+$00.700 and G41.16$-$00.20 the SED fits are poorly constrained
because of the lack of far-infrared
($>$$450\mu$m) flux density measurements. The stellar mass, temperature, radius, total 
luminosity
of the central star, and other 
properties can be estimated. We found that these 
parameters for our targets are typical for massive YSOs (De Buizer et
al.\,\cite{debuizer12}; Pandian et al.\,\cite{pandian10}).
Here we confine the discussion to the total luminosity, $L_{tot}$, 
which depends on the inclination angle, i, of the model. The range of i
is listed in Table~\ref{source-prop}. 
 We do not include the other properties in the discussion as was done in De Buizer et
al.\,(\cite{debuizer12}) since they are probably inconclusive. The median,
minimum, and 
maximum values of $L_{tot}$ from the ten best SED fits based on their $\chi^2$ values 
are calculated for each of the 11 objects (Table~\ref{source-prop}).

\begin{table*}
\caption{Colors and luminosities of MIR counterparts of methanol and non-methanol sources. 
 The W49N region is omitted because of its
methanol maser emission complexity and the lack of IR data at sufficient
resolution.}
\label{table_EGO}       
\begin{tabular}{lllccccccc}
\hline\hline
Name & l     & b     &1 rms             &  LSR Velocity range & [3.6]-[4.5] & [4.5]-[5.8] &  \multicolumn{3}{c}{Luminosity
($L_{\sun}$) at}\\
     & (deg) & (deg) & (mJy)            &    (km\,s$^{-1}$) &            &             &   3.6$\mu m$ & 4.5$\mu m$ &   5.8$\mu m$  \\
\hline
{\bf G037.7534-00.1892} & {\bf 37.753421} & {\bf -0.189203} &   20 & 18; 108 & 2.21 & 1.43 & 0.91 & 3.82 & 7.88 \\
G037.7540-00.1853       &  37.754001      & -0.185311       &  80  & 18; 108 & 2.55 & 1.52 & 0.26 & 1.50 & 3.35 \\
G037.7541-00.1874       &  37.754118      & -0.187425       &  80  & 18; 108 & 1.51 & -    & 0.21 & 0.46 & -    \\ 
G037.7510-00.1906       &  37.750999      & -0.190664       &  80  & 18; 108 & 0.88 & 2.90 & 0.68 & 0.83 & 6.66 \\ 
G037.7632-00.2150       &  37.763260      & -0.215031       &  90  & 18; 108 & 2.16 & 1.13 &10.56 &41.84 &65.51 \\
{\bf G040.2819-00.2197} & {\bf 40.281918} & {\bf -0.219794} &  10  & 18; 108 & 1.21 & 0.36 & 3.86 &28.47 &21.87 \\
{\bf G040.4249+00.7000} & {\bf 40.424941} & {\bf  0.700050} &  22  &$-$30; 60& 1.49 & 1.98 & 7.27 &15.55 &53.02 \\
G040.4308+00.7062       &  40.430894      &  0.706228       &  90  &$-$30; 60& -    & -    & -    & 8.45 &-     \\
G040.4184+00.7039       &  40.418421      &  0.703903       &  80  &$-$30; 60& 1.96 & 0.97 & 0.20 & 0.68 & 0.91 \\
{\bf G041.1232-00.2203} & {\bf 41.123224} & {\bf -0.220356} &  15  & 18; 108 & 1.19 & -    & 0.85 & 1.38 &-     \\
{\bf G041.1566-00.1968} & {\bf 41.156641} & {\bf -0.196881} &  10  & 18; 108 & 1.84 & -    & 0.23 & 0.67 &-     \\
G041.1580-00.1941       &  41.158047      & -0.194100       &  70  & 18; 108 & 0.26 & 1.50 & 4.59 & 3.17 & 6.94 \\
G041.1565-00.1985       &  41.156513      & -0.198536       &  70  & 18; 108 & 1.73 & -    & 0.13 & 0.35 &-     \\
G041.1560-00.2010       &  41.156050      & -0.201089       &  70  & 18; 108 & 2.04 & 0.39 & 0.13 & 0.48 & 0.38 \\
{\bf G041.2261-00.1970} & {\bf 41.226184} & {\bf -0.197014} &  15  & 18; 108 & 1.15 & -    & 1.84 & 2.87 &-     \\
{\bf G041.3475-00.1364} & {\bf 41.347554} & {\bf -0.136439} &  10  &$-$30; 60& -    & 1.30 & -    & 3.54 & 6.46 \\
{\bf G045.4671+00.0530}$^{a}$ & {\bf 45.467151} & {\bf  0.053064} &  7  & 18; 108 & 2.41 & 1.56 & 1.60 & 8.01 &18.48 \\
G045.4661+00.0457       &  45.466154      &  0.045739       &   60  & 18; 108 & -    & -    & -    & -    & 9.11 \\
G045.4692+00.0511       &  45.469271      &  0.051114       &   60  & 18; 108 & 1.16 & 0.77 & 8.73 &13.80 &15.47 \\
{\bf G045.4725+00.1335}$^{b}$ & {\bf 45.472562} & {\bf  0.133497} &  17  & 18; 108 & -    & 1.30 & -    &13.90 &25.48 \\
G045.4699+00.1327       &  45.469912      &  0.132772       &   70  & 18; 108 & -    & 2.11 & -    & 1.12 & 4.30 \\
{\bf G045.4925+00.1257} & {\bf 45.492565} & {\bf  0.125750} &   10  & 18; 108 & 2.23 & 0.19 & 0.26 & 1.10 & 0.72 \\
{\bf G059.7828+00.0647} & {\bf 59.782861} & {\bf  0.064733} &   6  &$-$30; 60& 1.85 &$-$0.07&54.14&161.88&83.85 \\
\hline
\end{tabular}
\tablefoot{The {\bf MIR} objects showing maser emission are in bold. Weak and diffuse 4.5$\mu$m emission also appears 
from $^{(a)}$ G45.4547+00.05986 and $^{(b)}$ G45.47658+00.13171 and
G45.46791+00.13492, but MIR flux densities are not listed in the GLIMPSE catalog
owing to very extended and complex morphology.} 
\end{table*}

\begin{table*}
\caption{Sources properties. }
\label{source-prop}       
\centering            
\begin{tabular}{lllccrc}
\hline\hline
Source & D$_{\rm near}$/D$_{\rm far}$ & D$_{\rm adopted}$ &  $L_{\rm tot}$&
i& L$_{\rm m(EVN)}$ & L$_{\rm m(EVN)}$/ L$_{\rm m(32m)}$ \\
   & (kpc) & (kpc) & ($10^{3}L_{\sun}$) & ($^{\rm o}$)& ($10^{-6} L_{\sun}$) &  \\
\hline
G37.753$-$00.189 & 3.5/9.8 & 8.8$^{\rm a}$ & 9.06$^{+5.70}_{-0.94}$ &32--57&  0.13  & 0.14 \\
G40.282$-$00.219 & 5.1/7.8 & 4.9$^{\rm b}$ & 94.8$^{+121.2}_{-45.4}$ &32--57& 2.47  & 0.36 \\ 
G40.425$+$00.700 & 1.3/11.5 & 11.4$^{\rm a}$  & 10.06$^{+11.35}_{-2.75}\uparrow$ &32-87&  16.64 & 0.76 \\
G41.123$-$00.220 & 4.2/8.4 & 8.7$^{\rm b}$ & 7.19$^{+8.11}_{-4.67}$  &32--49&   0.55  &      \\ 
G41.16$-$00.20   & 4.1/8.5 & 8.7$^{\rm b}$ & 0.23$^{+0.12}_{-0.10}\uparrow$&41--81&   0.14  &      \\ 
G41.226$-$00.197 & 3.8/8.8 & 8.7$^{\rm b}$ & 1.80$^{+29.70}_{-1.09}$ &18--32&   2.22  & 0.42 \\ 
G41.348$-$00.136 & 1.0/11.6 & 11.6$^{\rm b}$ & 290$^{+196}_{-189}$ &32--87&   10.63 & 0.62 \\  
G43.165$+$00.013 & --/11.8 & 11.11$^{\rm c}$  &  &&   9.28 &      \\
G43.171$+$00.004 & --/11.8 & 11.11$^{\rm c}$  &  &&   1.47 &      \\ 
G43.167$-$00.004 & --/11.8 & 11.11$^{\rm c}$  &  &&   0.07 &      \\
G43.149$+$00.013 & --/11.8 & 11.11$^{\rm c}$  &  &&   0.46 &      \\
W49N total       &  &  &           && 11.28 & 0.21 \\ 
G45.467$+$00.053 & 4.2/7.6 & 7.2$^{\rm b}$ & 143$^{+189}_{-0}$ &32--70&   1.41 & 0.57 \\  
G45.473$+$00.134 & 5.9/5.9 & 6.9$^{\rm b}$ & 28.3$^{+29.30}_{-15.00}$ &32--87&  1.05 & 0.35 \\ 
G45.493$+$00.126 & 4.3/7.5 & 7.1$^{\rm b}$ &  18.70$^{+116.30}_{-5.20}$ &32--41&  0.78 & 0.43 \\ 
G59.782$+$00.065 & 3.5/5.0 & 2.2$^{\rm d}$ &  7.65$^{+1.41}_{-2.13}$ &18--41&  1.78 & 0.71 \\
\hline
\end{tabular}
\tablefoot{ Kinematic distance ambiguity resolved from 
          $^{\rm (a)}$ Watson et al.\,(\cite{watson03}), $^{\rm (b)}$ Pandian et al.\,(\cite{pandian09}); trigonometric distance from 
          $^{\rm (c)}$ Zhang et al.\,(\cite{zhang13}), $^{\rm (d)}$ Xu et al.\,(\cite{xu09}). Lower limit of $L_{\rm tot}$
(marked by $\uparrow$) is given because the SED fits were poorly
constrained (Fig.~\ref{sed-on}).
} 
\end{table*}  

\begin{figure}[b]
\centering
   \includegraphics{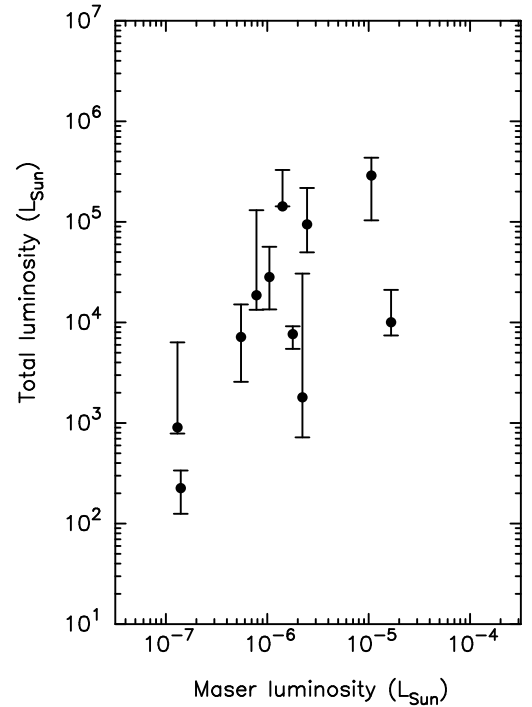}
    \caption{The relationship between the methanol maser luminosity as
obtained from the EVN observations and the total luminosity of the central star. The median values
    of the total luminosity are shown and the vertical bars show the range of the total luminosity.} 
    \label{total_mL}
    \end{figure}

We searched for relationships between the total luminosity of the star 
and the observed properties of the maser emission.
The maser (isotropic) luminosity of each target is listed in
Table~\ref{source-prop} as L$_{\rm m(EVN)}$. We also give the ratio of the
maser luminosity obtained from the EVN observations and single-dish data
(L$_{\rm m(EVN)}$/L$_{\rm m(32m)}$). The average maser luminosity is
(3.27$\pm$1.28)$\times$10$^{-6}$L$_{\sun}$ 
and the median value is 1.41$\times$10$^{-6}$L$_{\sun}$ (for all 15
targets). These values from the 32~m dish are (14.40$\pm$7.85)$\times$10$^{-6}$L$_{\sun}$ and
4.14$\times$10$^{-6}$L$_{\sun}$, respectively (for 10 sources). 

One can see that the isotropic maser luminosity and the total luminosity show some
correlation (Fig.\,\ref{total_mL}). However, we failed to obtain a consistent
fit indicating poor statistics (the correlation coefficient shows a very significant 
uncertainty of 150\%). A larger sample (including more distant
objects to avoid a bias due to the distance) is needed to verify if the
higher maser luminosity is related to higher stellar luminosity. 
We agree that this relation could put some constraints
 on the pumping mechanism or indicate that the larger clump/core just
 provides a longer maser amplification column as pointed out
by Urquhart et al.~(\cite{urquhart13}) for analysis of the isotropic maser
luminosities and masses of maser-associated submillimetre continuum
sources. More luminous central sources 
 would excite larger regions around them and a maser amplification
paths would be longer.

\subsection{MIR colors and luminosities}
\begin{figure}[h]
\centering
   \includegraphics{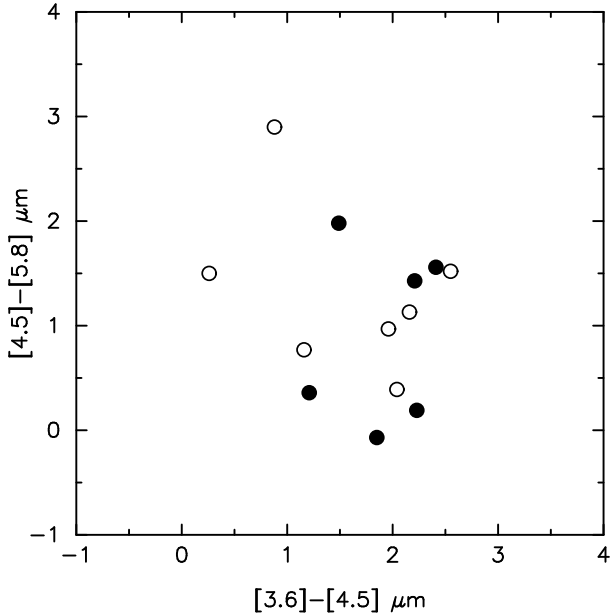}
    \caption{The color--color diagram of MIR counterparts with methanol
emission (black circles) and without
(open circles) as listed in Table~\ref{table_EGO}.} 
    \label{colors}
\end{figure}

The IRAC colors of objects listed in Table~\ref{table_EGO} do not show statistically 
significant differences between maser and non-maser sources for our sample 
(Fig.~\ref{colors}). This appears to be consistent with the survey of 
6.7\,GHz methanol masers in a sample of 20 MYSO outflow candidates
(the EGOs) (Cyganowski et al.\,\cite{cyganowski09}), where the detection rate was about 64\%.
Ellingsen~(\cite{ellingsen07}) selected a large sample (200) of GLIMPSE
sources, likely to be HMSFRs, on the basis of their MIR
colors (either bright at 8.0~$\mu$m or with extreme [3.6]--[4.5]
colors, and devoid of known methanol maser emission) and obtained a 
6.7~GHz maser detection rate of about 19\%.

The luminosity at 4.5$\mu$m is calculated with the following equation 
$L_{\nu}(4.5\mu m)[L_{\sun}] = 4.706 \times 10^{-3} D^{2}[\rm kpc]$ $S_{\nu}(4.5\mu m)[\rm mJy]$. 
 The flux is taken from the NASA/IPAC Infrared Science Archive via Gator
(Sect.~3). Similar equations are used
to calculate the luminosities at 3.6$\mu$m and 5.8$\mu$m. It is assumed that the non-maser objects in the field  
centered at the maser source are members of the same cluster of massive young stellar objects.  
Median values of $L_{\nu}(4.5\mu m)$ for maser and non-maser MIR objects are
3.82$L_{\sun}$ and
1.12$L_{\sun}$, respectively. The same values at IRAC bands 3.6$\mu$m and
5.8$\mu$m are 1.6 (maser MIR), 0.26 (non-maser MIR) and 20.18 (maser
MIR), 6.66$L_{\sun}$ (non-maser MIR), respectively.
We conclude that the likelihood of maser occurrence increases for IR bright objects. This supports a view 
that the appearance of the maser emission is dependent on the energy output of the central object. 
Analysis of a larger sample is needed to confirm our result. Moreover,
observations at sufficient angular resolution at other wavelengths would help to
distinguish whether that effect is related to the evolutionary stage of the
central object. 

\subsection{Maser emission characteristics and origin}
There appears to be a correlation of the extent of the maser region with its velocity
range (Fig.~\ref{size-vel-lum}). To obtain more
reliable statistics we added measurements from 31 targets from our previous
sample (Bartkiewicz et al.~\cite{bartkiewicz09}). We
confirm, as was reported by Pandian et al.~(\cite{pandian11}),  
that the larger the emission extent of a maser, the wider its velocity
width (a log-log correlation). 
However, no dependence is clear from the second relation, maser
luminosity vs. emission extent (Fig.~\ref{size-vel-lum}). In the plot we
highlight seven masers with clear ring-like
morphology (six from Bartkiewicz et al.~\cite{bartkiewicz09} that
consist of more than four groups of maser spots, the obvious fitting of an
ellipse, and the G40.425$+$00.700 source from this sample). They all tend to fall in
a region with higher emission extent ($>$330~AU) and velocity width
($>$4.8~km~s$^{-1}$). The ring-like 6.7~GHz methanol maser discovered
toward Cep~A has similar properties with the extent of 960~AU and 
$\Delta$V=3.2~km~s$^{-1}$ (Sugiyama et al.~\cite{sugiyama08}, Torstensson et al.~\cite{torstensson11}).
 Three new methanol maser rings (with more than four groups of maser
spots) have been detected recently by Fujisawa et al.~(\cite{fujisawa14}) 
toward 000.54$-$00.85 SE, 002.53$+$00.19, and 025.82$-$00.17. They also showed the high 
linear extent (1500--4500~AU) 
and significant velocity width ($\Delta$V from 8.8~km~s$^{-1}$ to 16~km~s$^{-1}$).

\begin{figure}[h]
\centering
   \includegraphics{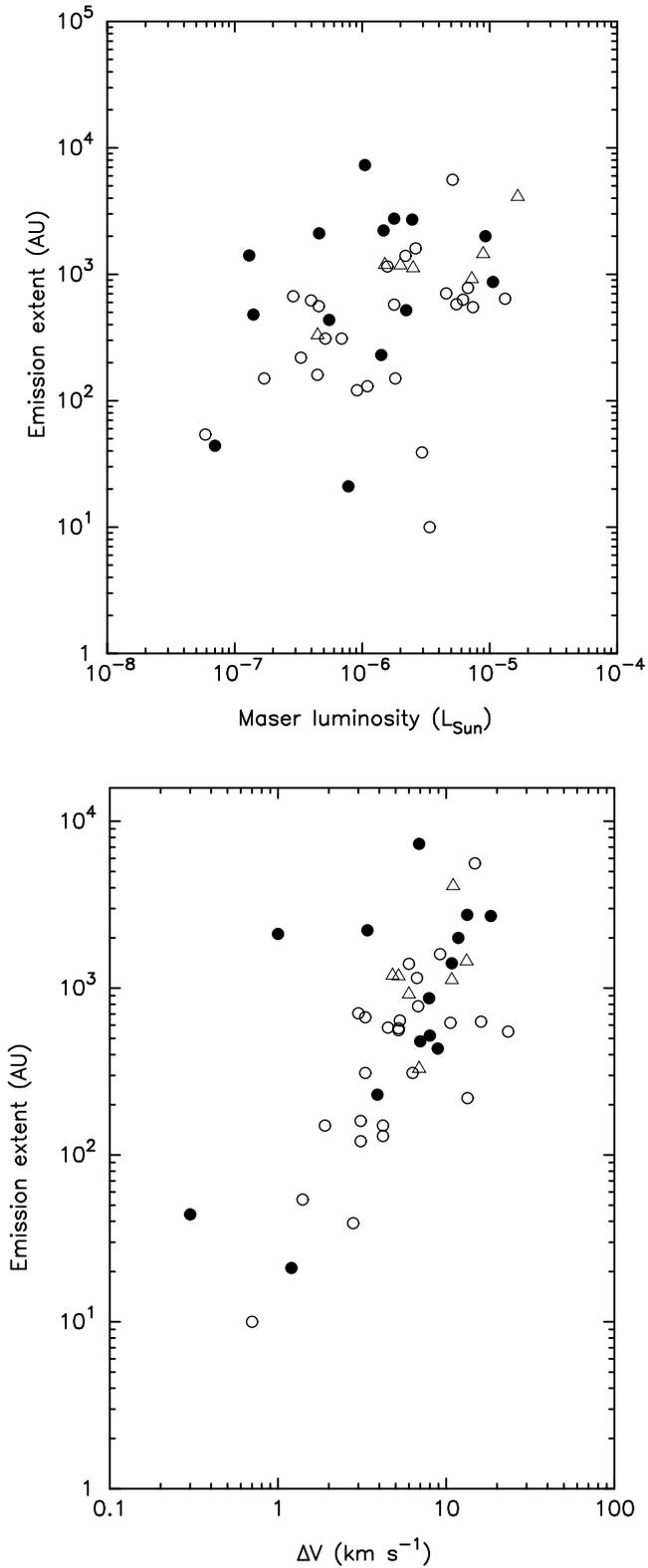}
    \caption{{\bf Top:} The relationships between the isotropic maser 
luminosity as obtained from the EVN data and its spatial extent (the major axis) and {\bf Bottom:} between
 the velocity extent of the methanol maser line and the spatial extent. The
targets from this paper are marked with black circles; the targets from Bartkiewicz et
al.~(\cite{bartkiewicz09}) are marked with open circles and triangles (the sources with clear
ring-like morphology containing more than four maser spot groups).} 
    \label{size-vel-lum}
    \end{figure}

Comparison of the methanol maser profiles taken from the EVN and single-dish observations reveals that about 57\% 
of the maser flux density is resolved out (Table \ref{source-prop}). The fraction of missing flux ranges from 24\% 
in G40.425$+$00.700 to 86\% in G37.753$-$00.189. Its value clearly does not depend on the distance of the source which 
implies that the occurrence of weak and diffuse maser emission is specific to an individual source.
The lack of information on the distribution of weak and diffuse emission may prevent
the determination of the full extent
and morphology of the maser source. We note that 12 sources in the sample were observed with the
EVLA and/or MERLIN with beamsizes of 0\farcs45$\times$0\farcs26 and
0\farcs060$\times$0\farcs035, respectively (Pandian et al.\,\cite{pandian10}). The overall
morphologies of these sources obtained with the VLBI agree well with those observed with lower angular resolution. 
This implies that almost all maser clouds have compact cores and the source structures are
actually recovered.

All the maser targets are usually associated with the brightest MYSOs within
each cluster showing an excess of
extended 4.5$\mu$m emission. However, only two of the eleven maser sources
(Table\,\ref{table_EGO}) are associated with
``possible'' MYSO outflow candidates in the EGO catalog (Cyganowski et
al.\,\cite{cyganowski08}). We note that these   
sources G40.2819$-$0.2197 and G45.4725$+$0.1335 are in a group of objects with
the linear projected size of maser
emission greater than 2700\,AU (Table\,\ref{table2}). Therefore, the
maser emission may arise in outflows
or in multiple nearby point sources. The source G45.4661$+$0.0457 reported by
Cyganowski et al. (\cite{cyganowski08})
as a ``likely'' MYSO outflow candidate is a non-maser source. They estimated a
lower limit of detection rate of 6.7\,GHz
maser emission in ``likely'' MYSO outflow candidate EGOs as high as 73\%.
We can conclude that in our sample the maser emission is rarely associated
with outflows traced by extended
4.5$\mu$m emission. Most of our maser sources are associated with the MIR
objects easily identified via color selection
in the GLIMPSE I Spring ’07 Archive (Table\,\ref{table_EGO}). The appearance
of their 4.5$\mu$m emission excess of angular
extents from a few to less than 10\arcsec \, could be related to high
extinction (Rathborne et al.~\cite{rathborne05}; Cyganowski et
al.~\cite{cyganowski08}).
This implies that most of our masers are associated with highly embedded
MYSOs.
Coarse angular resolution MIR images cannot be easily compared with the VLBI
images of mas resolution to examine whether
the maser arises in outflows or in disks. It seems that the proper motion
studies of maser components is only one way    
to verify the ongoing phenomenon.

\section{Conclusions}
We successfully imaged the 6.7~GHz methanol emission toward 15 
targets (four of them belong to the complex HMSFR W49N). Using
a short correlation time (0.25~s) we were able to image significantly
offset emission of $-$54'' in RA and $-$56'' in Dec (G43.149$+$00.013) from
the pointing center. Although the VLBI resolves some of the emission, there is no problem studying the morphology; almost all maser clouds must have compact
cores. The maser emission was always associated with the strongest MIR counterpart in
the clusters, so we conclude that the appearance of the maser emission is
related to the IR brightness of the central MYSO. The maser (isotropic)
luminosity and the total luminosity of the central object are likely
to correlate; however, a larger sample is needed to verify this relation. 
The maser linear extent is related to its spectrum width. We also note that the
spectra of methanol rings are the widest among the sample and their emission
also appears more extended on the sky.

\begin{acknowledgements}
AB and MS acknowledge support from the National Science Centre Poland 
through grant 2011/03/B/ST9/00627. We thank Dr.~Jagadheep Pandian for useful
discussions. 
We thank the Referee for a detailed and constructive report which improved
this article.
This work has also been supported by the European Community
Framework Programme 7, Advanced Radio Astronomy in Europe, grant agreement
No.~227290. 
This research has made use of the VizieR catalog access tool,
CDS, Strasbourg, France. This research
made use of data products from the Midcourse Space Experiment. 
This research has also
made use of the NASA/ IPAC Infrared Science Archive at
http://irsa.ipac.caltech.edu/, which is operated by
the Jet Propulsion Laboratory, California Institute of Technology, under
contract with NASA. This paper
made use of information from the Red MSX Source survey database at
www.ast.leeds.ac.uk/RMS which was constructed with support from the Science
and Technology Facilities Council of the UK.

\end{acknowledgements}

\Online

\begin{appendix}
\section{Figures}

   \begin{figure*}
   \centering
    \includegraphics[scale=0.75]{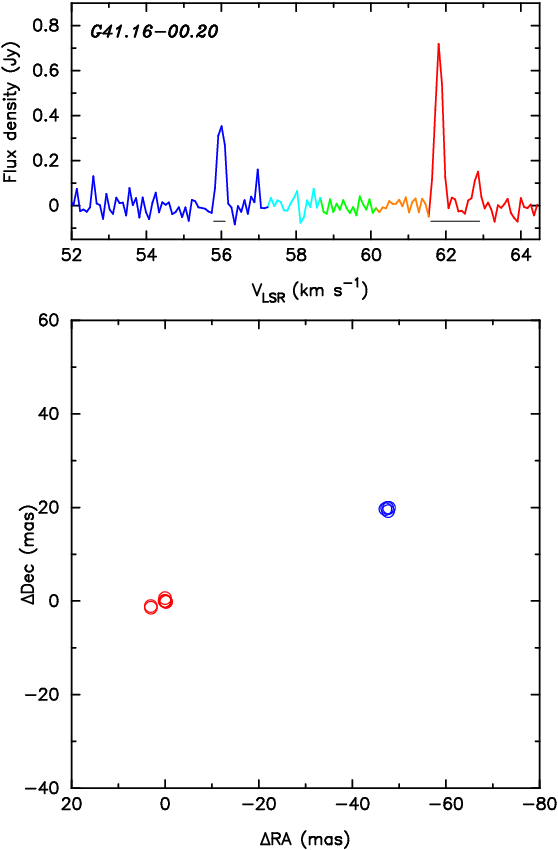}
    \includegraphics[scale=0.75]{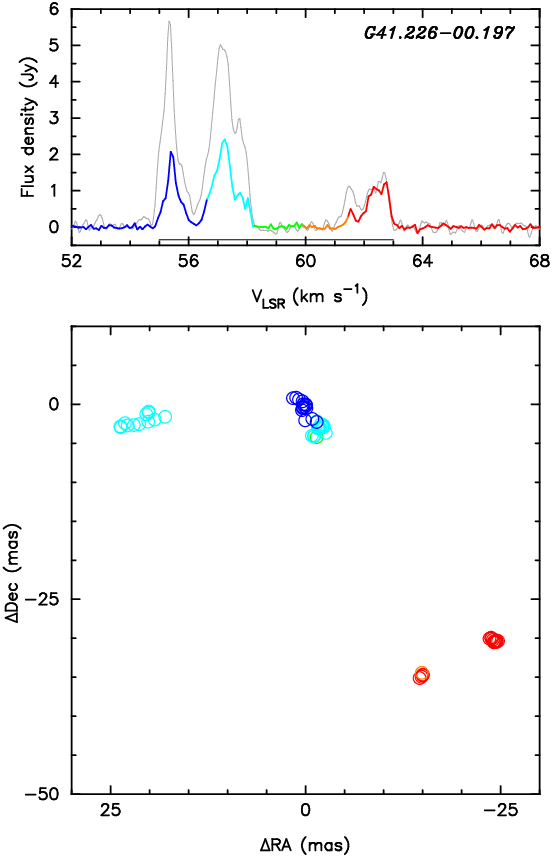}
    \includegraphics[scale=0.75]{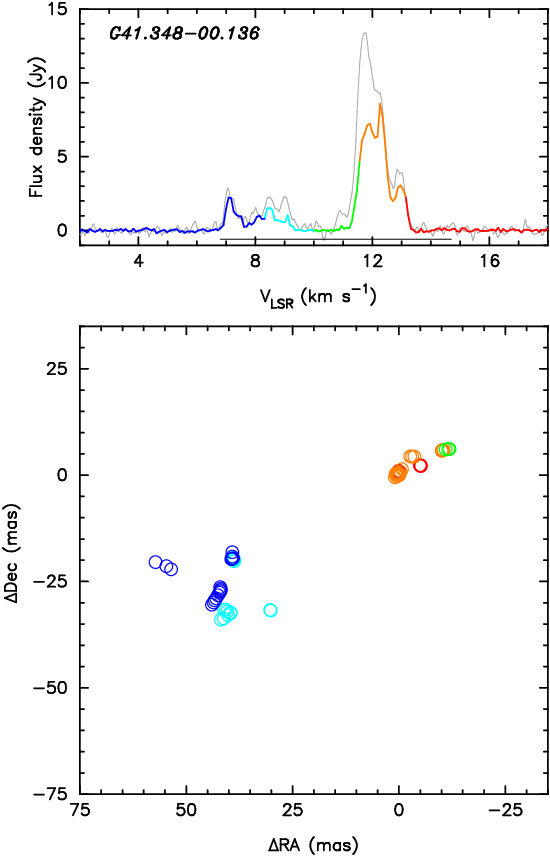}
    \includegraphics[scale=0.75]{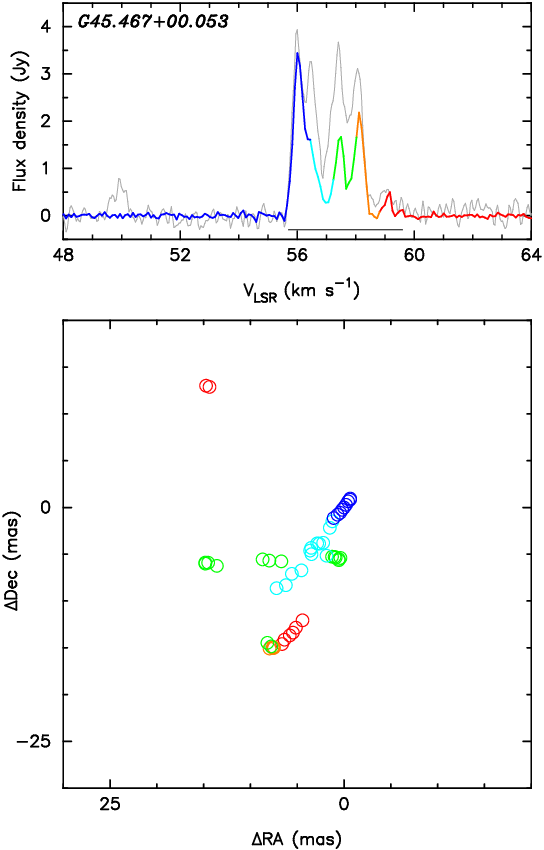}
    \caption{Spectra and maps of 6.7~GHz methanol masers detected using the EVN.
    The names are the Galactic coordinates of the brightest spots listed
    in Table~\ref{table2}. The colors of circles relate to the LSR velocities 
    as shown in the spectra. 
    The map origins are the locations of the brightest spots
    (Table~\ref{table2}). The gray lines show the Torun 32~m dish spectra.
    If needed, the separate scale of the flux density is presented on the
    left (EVN) and right (Torun) sides. The thin bars under the spectra
    show the LSR velocity ranges of spots displayed. The plots for the
    first four targets are presented in Fig.~\ref{distrib}.} 
\addtocounter{figure}{-1}
    \end{figure*}

   \begin{figure*}
   \centering
    \includegraphics[scale=0.75]{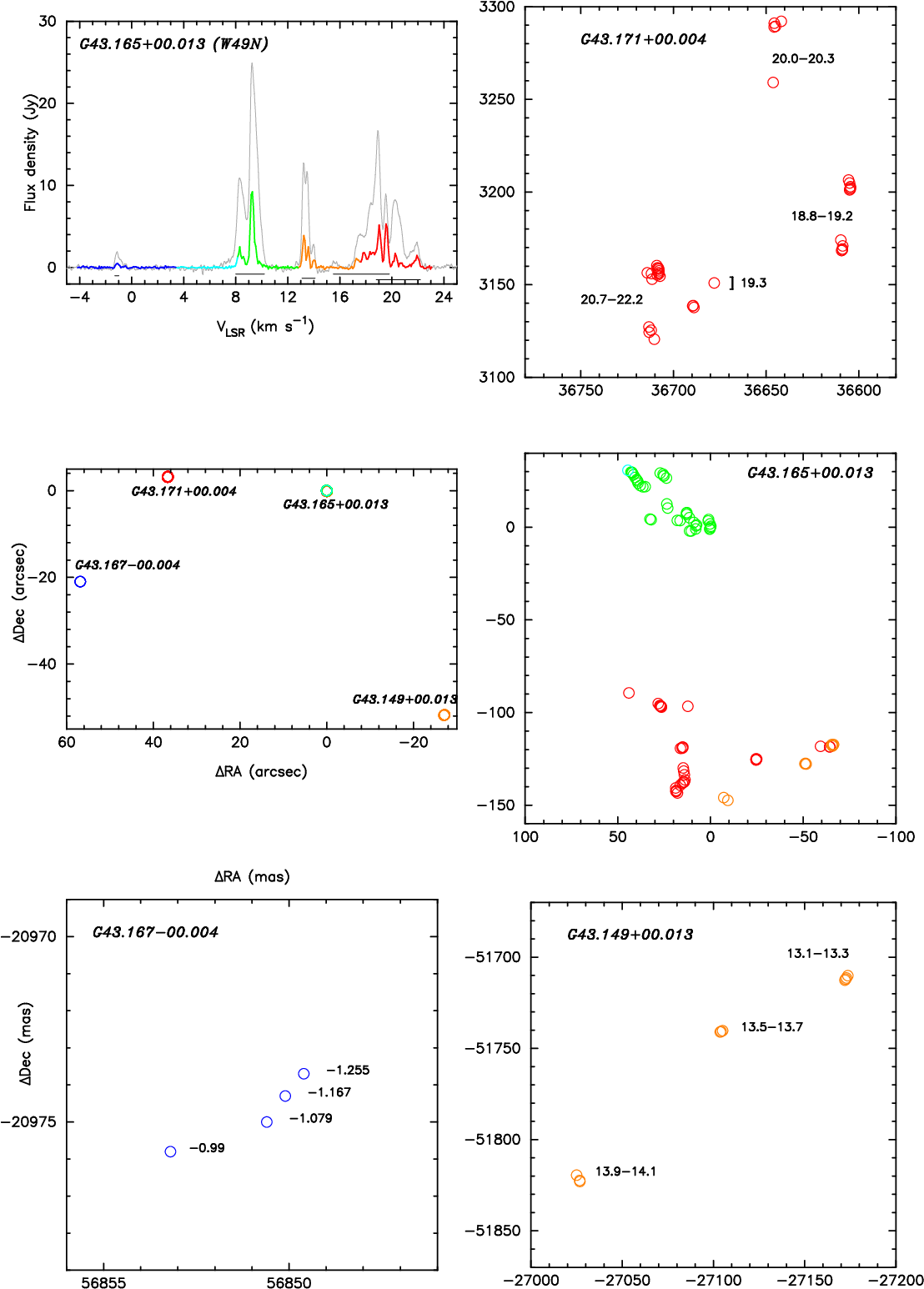}
    \caption{continued. For clarity, in G43.171$+$00.004, G43.167$-$0.004, and
G43.149$+$0.013, we listed the LSR velocities in km~s$^{-1}$ (or LSR velocity range) of a given maser
(or maser groups).} 
\addtocounter{figure}{-1} 
   \end{figure*}

   \begin{figure*}[b]
   \centering
    \includegraphics[scale=0.75]{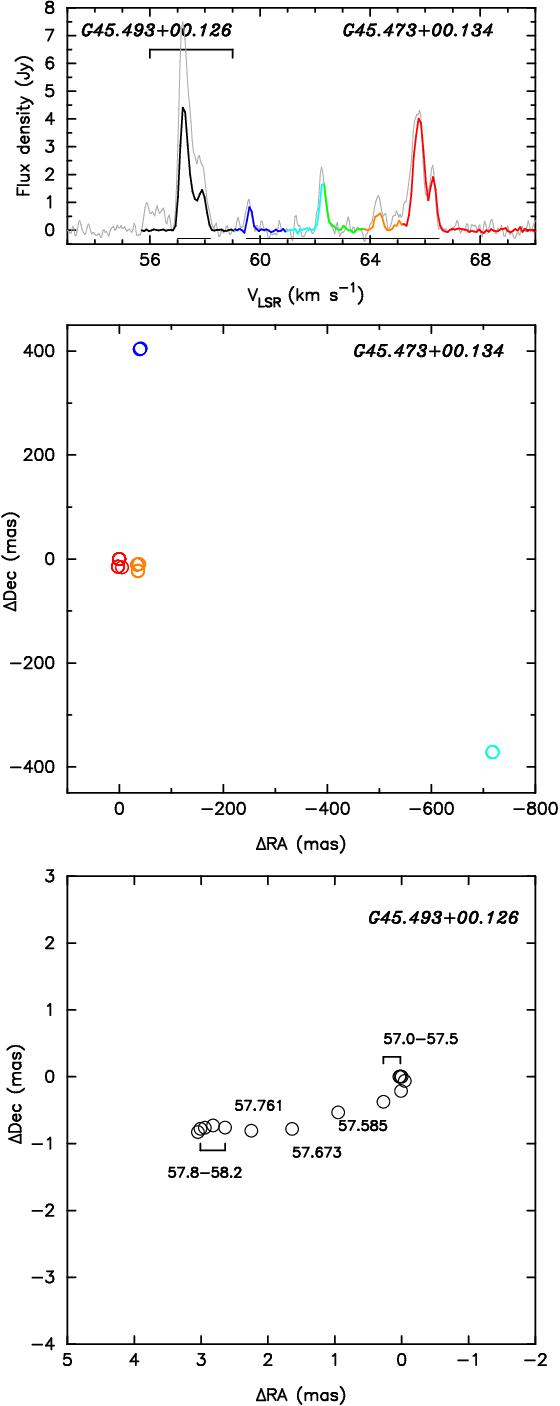}
    \includegraphics[scale=0.75]{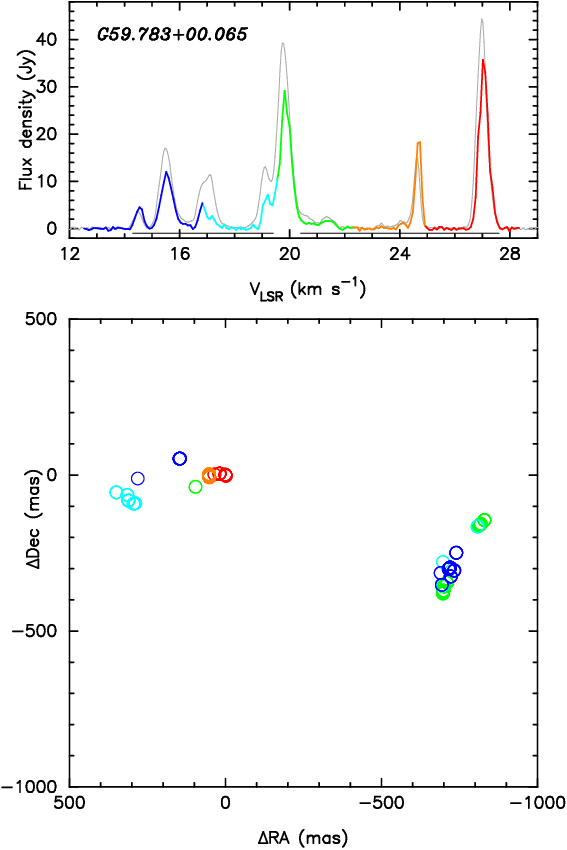}
    \caption{continued. In G45.493$+$00.126 we listed the LSR velocities in
    km~s$^{-1}$ (or LSR velocity range) of a given maser (or maser groups).} 
     \label{distrib2}
    \end{figure*}

   \begin{figure*}
    \centering
     \includegraphics[scale=1]{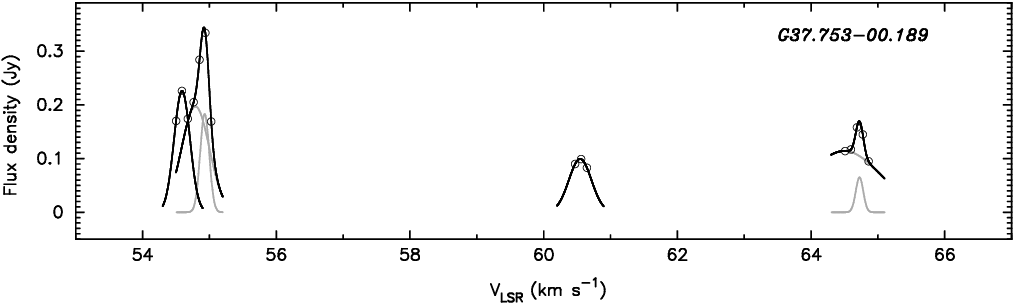}
     \includegraphics[scale=1]{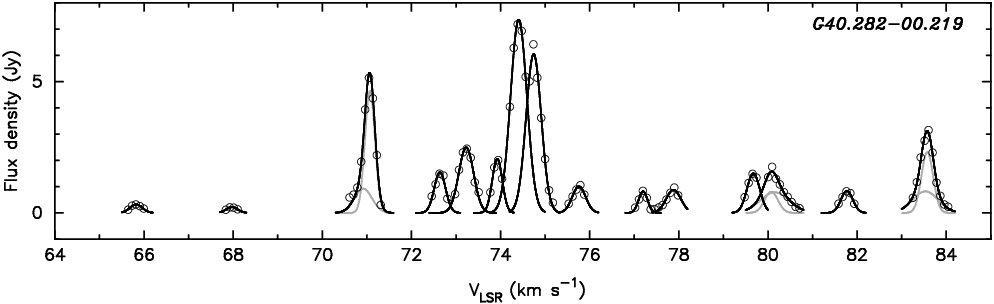}
     \includegraphics[scale=1]{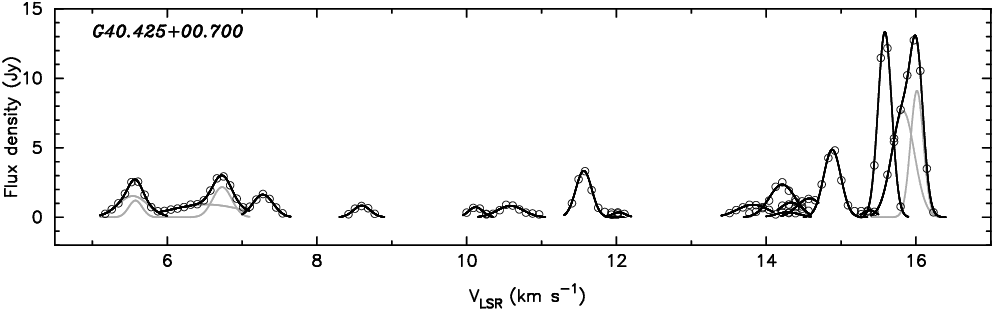}
     \includegraphics[scale=1]{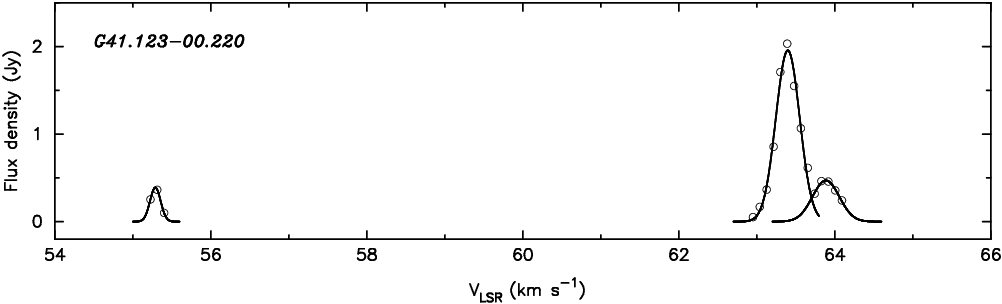}
    \caption{Spectra of individual 6.7~GHz maser clouds with Gaussian velocity
profiles. Each circle traces the emission level of a single maser spot as
presented in Figs.~\ref{distrib} and \ref{distrib2}. The
black line represents the fitting of a Gaussian function (or functions) as summarized 
in Table~\ref{table3}. The gray line presents the single Gaussian
fitting in a case of complex velocity profile of an individual component.}
\addtocounter{figure}{-1}
     \label{gauss}
   \end{figure*}
   \begin{figure*}
    \centering
     \includegraphics[scale=0.9]{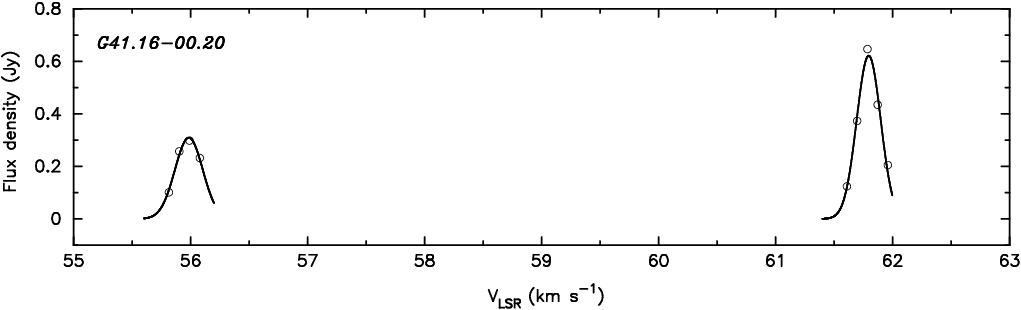}
     \includegraphics[scale=0.9]{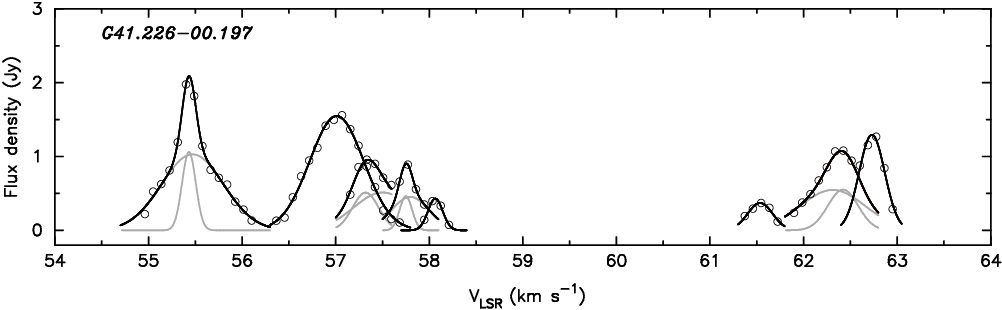}
     \includegraphics[scale=0.9]{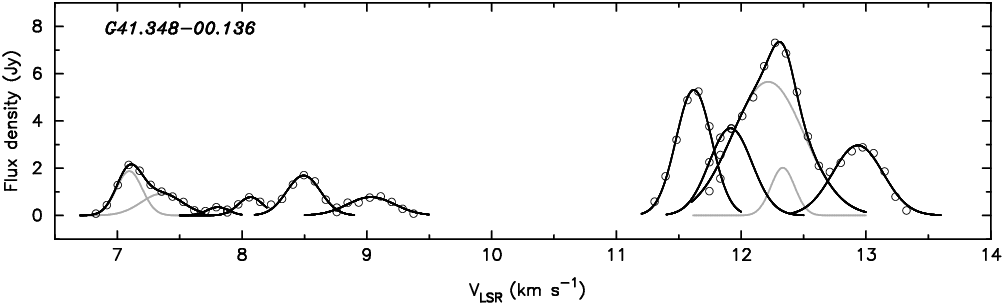}
     \includegraphics[scale=0.9]{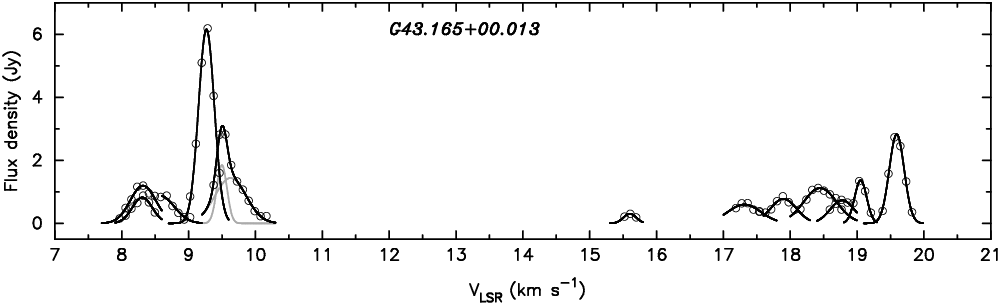}
     \includegraphics[scale=0.9]{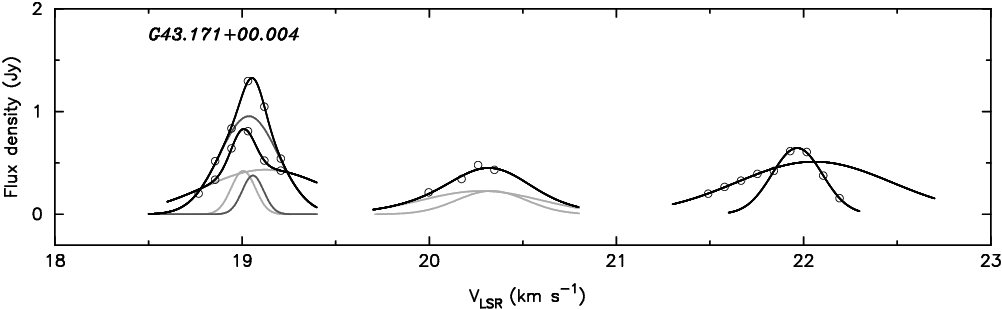}
     \caption{continued.}
\addtocounter{figure}{-1}
   \end{figure*}
   \begin{figure*}
    \centering
     \includegraphics[scale=1]{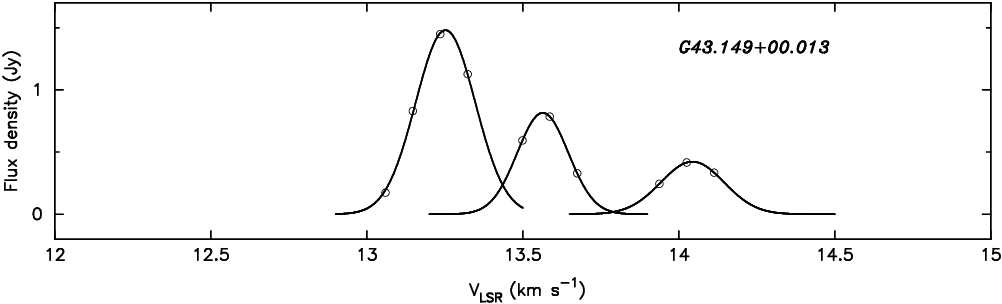}
     \includegraphics[scale=1]{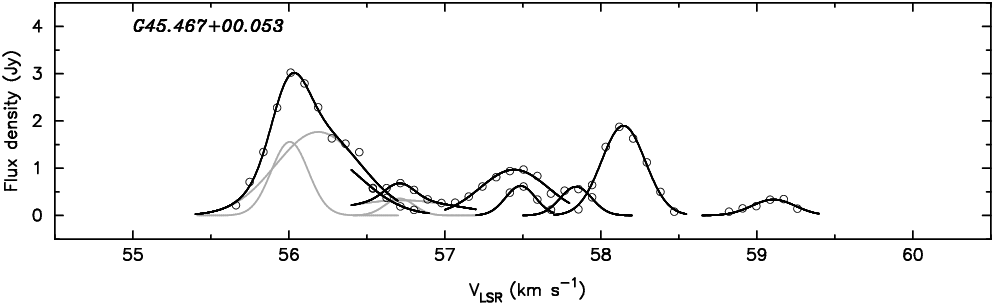}
     \includegraphics[scale=1]{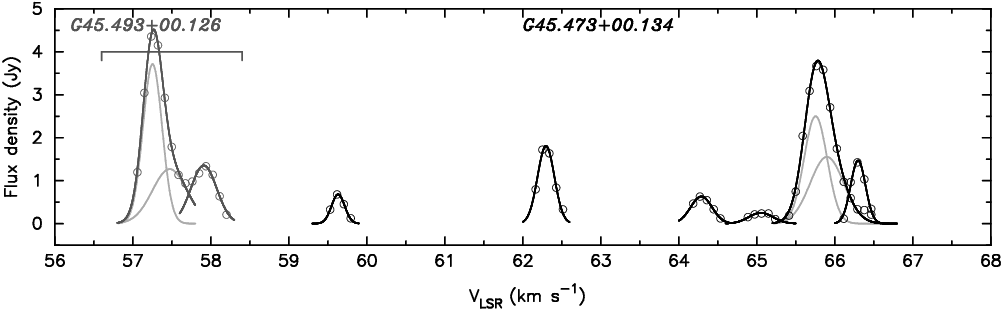}
     \includegraphics[scale=1]{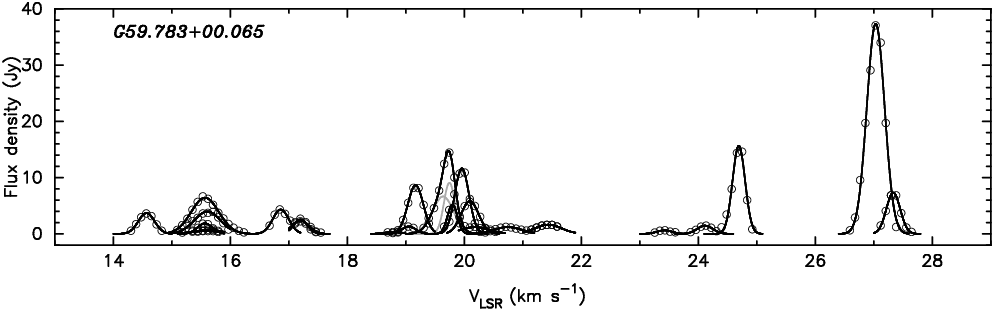}
     \caption{continued.}
   \end{figure*}

\begin{figure*}
\includegraphics[scale=0.75]{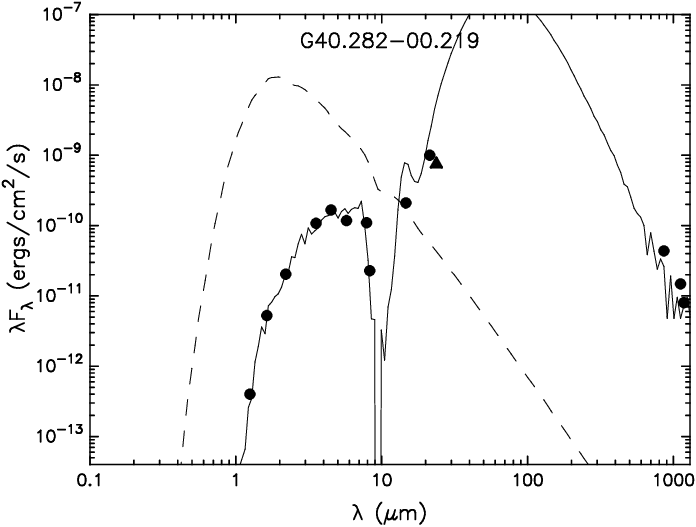}
\includegraphics[scale=0.75]{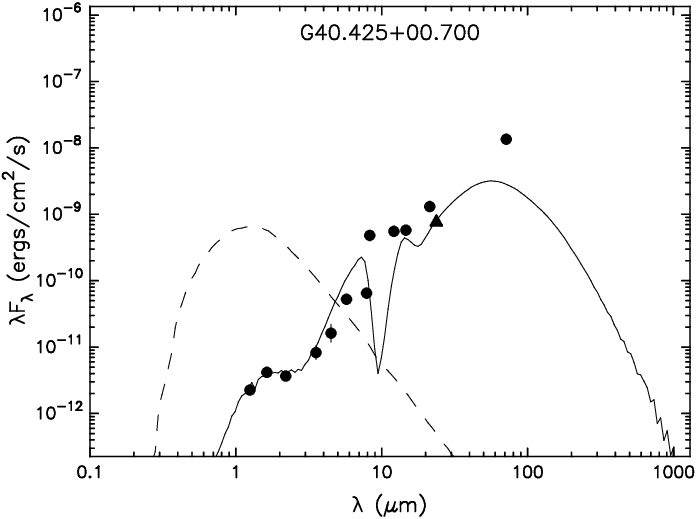}
\includegraphics[scale=0.75]{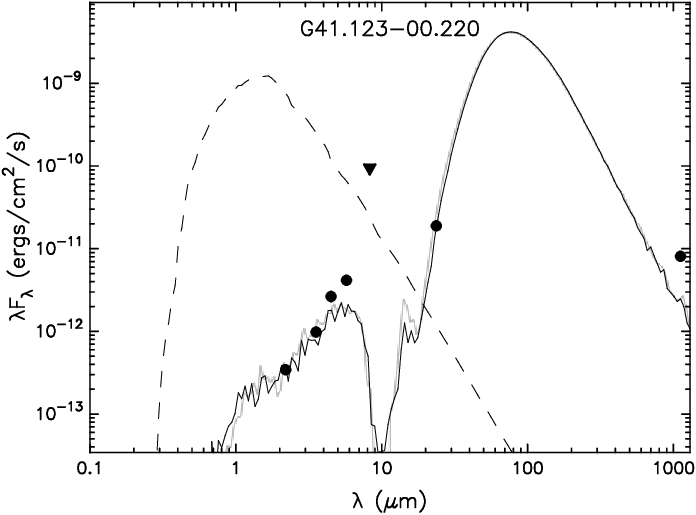}
\includegraphics[scale=0.75]{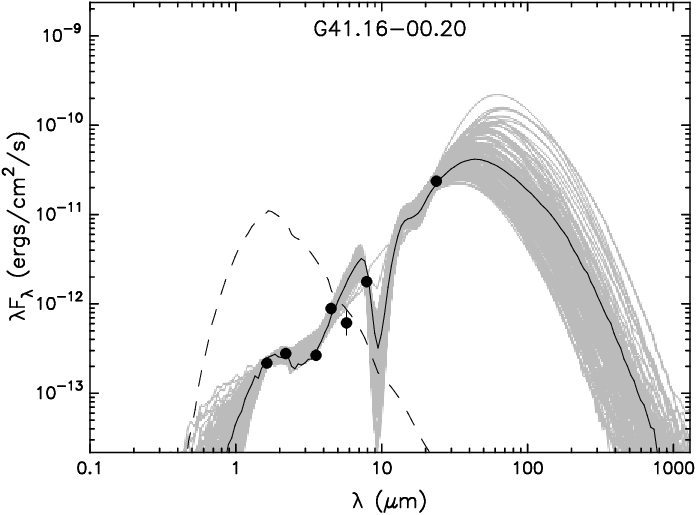}
\includegraphics[scale=0.75]{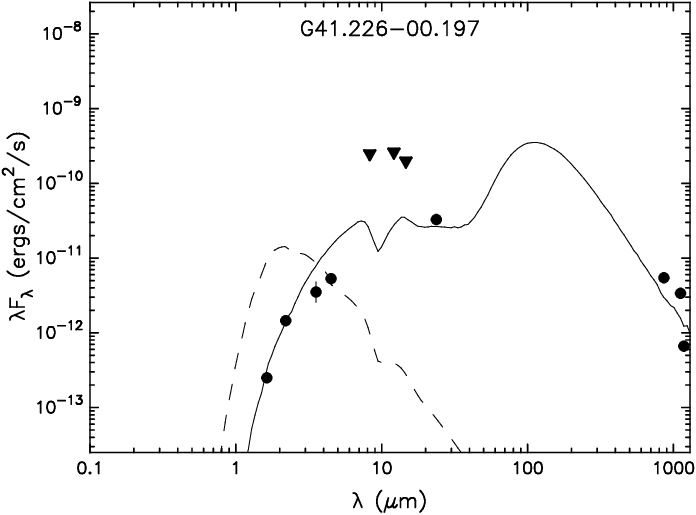}
\includegraphics[scale=0.75]{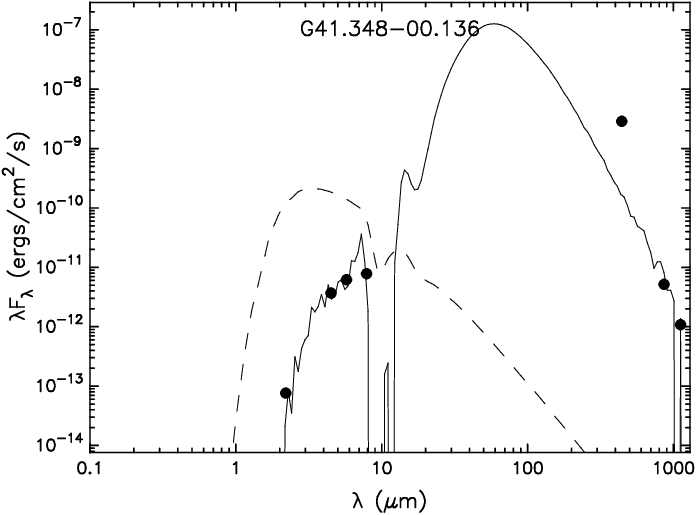}
 \caption{Fit to the spectral energy distribution for the remaining targets.}
 \label{sed-on} 
\addtocounter{figure}{-1}
\end{figure*}
\begin{figure*}
\includegraphics[scale=0.75]{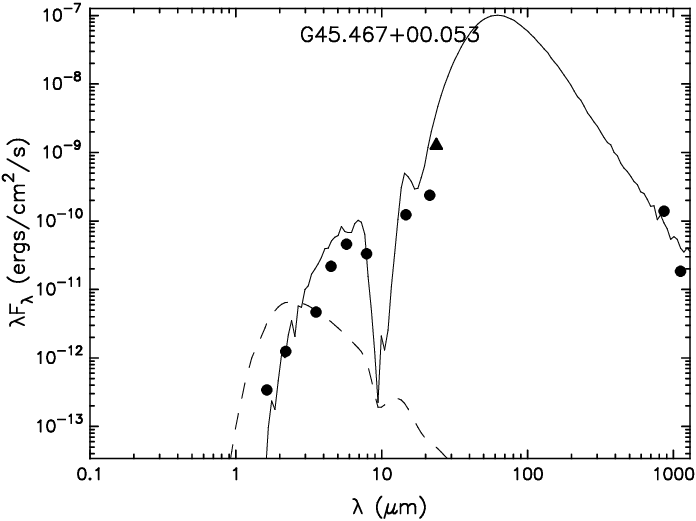}
\includegraphics[scale=0.75]{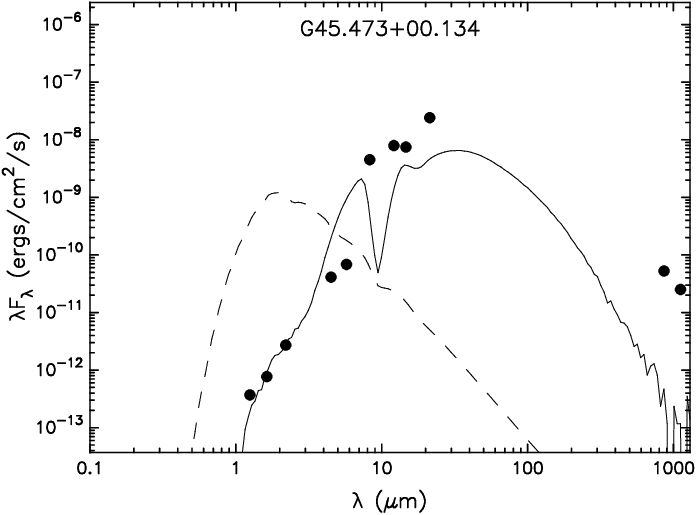}
\includegraphics[scale=0.75]{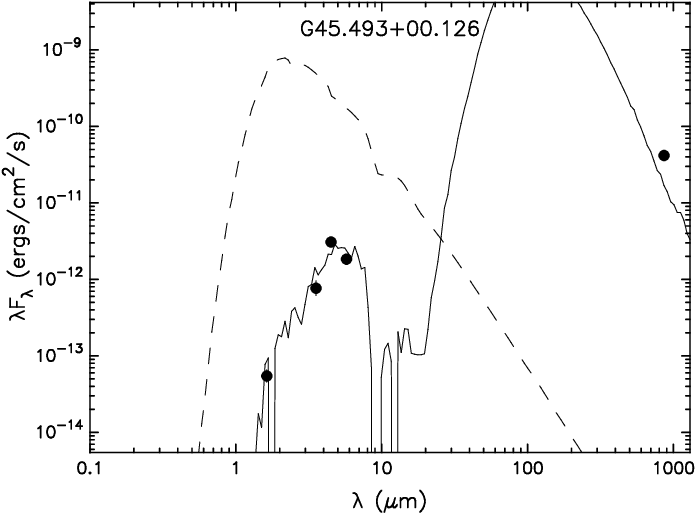}
\includegraphics[scale=0.75]{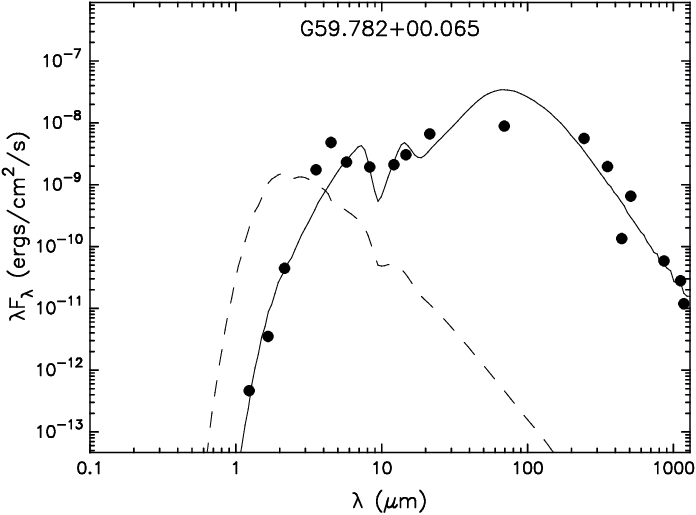}
 \caption{Continued.}
\end{figure*}
\end{appendix}

\begin{appendix}
\onecolumn
\section{Tables}
\begin{longtable}{c r c c c c c c ll}
\caption{\label{table3} Parameters of 6.7~GHz methanol maser clouds with Gaussian velocity profiles.
The coordinates are relative to the brightest spot of each source 
listed in Table~\ref{table2}. } \\       
\hline\hline
Cloud & $\Delta$RA & $\Delta$Dec & V$_{\rm p}$ & V$_{\rm fit}$ & FWHM & $S_{\rm p}$ & $S_{\rm fit}$ & 
$L_{\rm proj}^a$ & $V_{\rm grad}^a$\\
& (mas) & (mas) & (km~s$^{-1}$) & (km~s$^{-1}$) & (km~s$^{-1}$) & (Jy~beam$^{-1}$) & (Jy~beam$^{-1}$) &
(mas) & (km~s$^{-1}$~mas$^{-1}$)\\
&&&&&&&&(AU)&(km~s$^{-1}$~AU$^{-1}$)\\
\hline                    
\endfirsthead
\caption{Continued.}\\
\hline\hline
Cloud & $\Delta$RA & $\Delta$Dec & V$_{\rm p}$ & V$_{\rm fit}$ & FWHM & $S_{\rm p}$ & $S_{\rm fit}$ & $L_{\rm
proj}^a$ & $V_{\rm grad}^a$\\
& (mas) & (mas) & (km~s$^{-1}$) & (km~s$^{-1}$) & (km~s$^{-1}$) & (Jy~beam$^{-1}$) & (Jy~beam$^{-1}$) & 
(mas) & (km~s$^{-1}$~mas$^{-1}$)\\
&&&&&&&&(AU)&(km~s$^{-1}$~AU$^{-1}$)\\
\hline
\endhead
\hline
\endfoot
\hline
\endlastfoot
\multicolumn{10}{l}{\bf G37.753$-$00.189}\\
{\it 1}&  0.637 & -0.567 &54.59 &54.59 & 0.28 &   0.226 &  0.226 &  1.0(8.80) & 0.18(0.0205) \\ 
{\it 2}&  0.000 &  0.000 &54.94 &54.79 & 0.49 &   0.334 &  0.197 &  0.5(4.40) & --\\ 
&   &  &  &54.93 & 0.16 &    &  0.183 &   &  \\ 
{\it 3}& 82.047 & 45.336 &60.56 &60.55 & 0.40 &   0.099 &  0.099 & 1.1(9.68) & 0.17(0.01293) \\ 
{\it 4}& 136.529 & 61.645 &64.69 &64.50 & 1.30 &   0.158 &  0.114 & 1.3(11.44) &-- \\ 
&   &   &  &64.72 & 0.13 &    &  0.065 &   &  \\ 
\multicolumn{9}{l}{\bf G40.282$-$00.219}\\
{\it 1}& 304.688 &316.375 &65.82 &65.82 & 0.33 &   0.318 &  0.320 & 0.6(2.94) & 0.56(0.1142) \\ 
{\it 2}& 307.092 &318.456 &67.92 &67.97 & 0.30 &   0.210 &  0.215 & 1.1(5.39) &--\\ 
{\it 3}&  65.941 &459.945 &72.63 &72.64 & 0.30 &   1.490 &  1.545 & 2.9(14.21) & 0.12(0.0245) \\ 
{\it 4}&-114.287 &108.659 &71.05 &70.92 & 0.46 &   5.147 &  0.927 & 2.8(13.72) & 0.25(0.0510) \\ 
&   &   &  &71.07 & 0.26 &   &  4.644 &  &  \\ 
{\it 5}& 70.669 &440.315 &73.24 &73.23 & 0.39 &   2.447 &  2.495 & 4.7(23.03) & 0.11(0.0225) \\ 
{\it 6}&  2.195 &-10.457 &73.95 &73.93 & 0.26 &   2.009 &  2.054 & 3.0(14.70) & 0.09(0.0184) \\ 
{\it 7}&  0.000 &  0.000 &74.39 &74.41 & 0.43 &   7.189 &  7.355 &  1.5(7.35) & 0.23(0.0469) \\ 
{\it 8}& -1.940 & -0.055 &74.74 &74.70 & 0.16 &   6.423 &  1.765 & 5.4(26.46) & 0.10(0.0204) \\ 
&   &   &  &74.78 & 0.43 &    &  2.479 &   &  \\ 
&   &   &  &74.77 & 0.36 &    &  2.482 &   &  \\ 
{\it 9}& -1.769 & 21.829 &75.79 &75.76 & 0.35 &   1.058 &  1.007 & 1.1(5.39) & 0.31(0.0632) \\ 
{\it 10}&129.622 &120.422 &77.19 &77.20 & 0.24 &   0.836 &  0.817 & 8.0(39.20) & 0.05(0.0102) \\ 
{\it 11}&-38.592 & 24.791 &77.90 &77.88 & 0.39 &   1.007 &  0.877 & 2.4(11.76) & 0.18(0.0367) \\ 
{\it 12}&-57.163 & 92.929 &79.65 &79.67 & 0.33 &   1.485 &  0.750 & 0.9(4.41) & 0.30(0.0612) \\ 
 &  &  &  &79.67 & 0.35 &    &  0.751 &   &  \\ 
{\it 13}&-55.680 & 93.479 &80.09 &80.07 & 0.35 &   1.750 &  0.807 & 3.0(14.60) & 0.24(0.0490) \\ 
 &  &   &  &80.15 & 0.76 &    &  0.783 &   &  \\ 
{\it 14}& 34.607 &130.784 &81.76 &81.77 & 0.31 &   0.819 &  0.834 & 1.1(5.39) & 0.32(0.0653) \\ 
{\it 15}& 24.715 &142.690 &83.60 &83.58 & 0.31 &   3.152 &  2.312 & 2.4(11.76) & 0.38(0.0776) \\ 
 & &   &  &83.53 & 0.67 &   &  0.820 &   &  \\ 
\multicolumn{9}{l}{\bf G40.425$+$00.700}\\
{\it 1}&-133.685 &250.719 & 5.61 & 5.58 & 0.21 &   2.567 &  1.209 & 1.7(22.80) & 0.42(0.0368) \\ 
 &  &   &  & 5.54 & 0.47 &    &  1.533 &   &  \\ 
{\it 2}&-136.466 &252.412 & 6.75 & 6.52 & 1.11 &   2.940 &  0.924 & 2.5(28.50) & 0.42(0.0368) \\ 
 &  &  &  & 6.73 & 0.30 &    &  2.180 &  &  \\ 
{\it 3}&-138.216 &251.826 & 7.28 & 7.28 & 0.32 &   1.683 &  1.633 & 5.8(66.12) & 0.08(0.0070) \\ 
{\it 4}&-57.825 &312.915 & 8.60 & 8.60 & 0.26 &   0.819 &  0.841 & 2.5(28.50) & 0.14(0.0123) \\ 
{\it 5}&-176.304 & 84.229 &10.09 &10.12 & 0.22 &   0.678 &  0.731 & 3.5(39.90) & 0.07(0.0061) \\ 
{\it 6}&-168.356 &129.024 &10.53 &10.58 & 0.41 &   0.900 &  0.824 & 4.3(49.02) & 0.16(0.0140) \\ 
{\it 7}&-112.845 &255.440 &11.58 &11.56 & 0.25 &   3.317 &  3.340 & 5.6(63.84) & 0.11(0.0097) \\ 
{\it 8}&-119.122 &236.209 &12.02 &12.03 & 0.21 &   0.352 &  0.355 & 0.6(6.840) & 0.31(0.0272) \\
{\it 9}& -8.080 & -5.263 &13.78 &13.82 & 0.50 &   0.905 &  0.901 & 1.4(15.96) & 0.39(0.0342) \\ 
{\it 10}&-11.205 & -4.476 &14.39 &14.33 & 0.36 &   1.102 &  1.055 & 1.1(12.54) & 0.23(0.0202) \\ 
{\it 11}&-9.216 &-11.259 &14.22 &14.21 & 0.40 &   2.523 &  2.362 & 9.9(112.85) & 0.08(0.0070) \\ 
{\it 12}&-55.507 & 10.824 & 14.30 & 14.29 & 0.37 & 0.349 & 0.367 & 6.0(68.40) & 0.04(0.0035) \\
{\it 13}&  1.398 & -8.957 &14.57 &14.58 & 0.39 &   1.317 &  1.357 & 2.2(25.08) & 0.16(0.0140) \\ 
{\it 14}& -1.949 & -5.349 &14.92 &14.88 & 0.27 &   4.804 &  4.870 & 6.3(71.81)) & 0.10(0.0088) \\ 
{\it 15}& -7.565 & -3.551 &15.36 &15.38 & 0.21 &   0.480 &  0.502 & 1.4(15.96) & 0.13(0.0114) \\ 
{\it 16}&  6.805 & -6.417 &15.62 &15.59 & 0.21 &  12.172 & 13.340 & 1.2(13.67) & 0.29(0.0254) \\ 
{\it 17}&  0.000 &  0.000 &15.97 &15.83 & 0.36 &  12.742 &  7.650 & 1.0(11.40)) & 0.61(0.0535) \\ 
&  &   &  &16.01 & 0.20 &  &  9.115 &  &  \\ 
\multicolumn{9}{l}{\bf G41.123$-$00.220}\\
{\it 1}& 42.719 &-19.328 &55.31 &55.29 & 0.16 &   0.364 &  0.389 & 0.6(5.22) & 0.28(0.0322) \\ 
{\it 2}& -0.109 &  0.129 &63.30 &63.40 & 0.36 &   1.710 &  1.956 & 1.2(10.44) & 0.53(0.0609) \\ 
{\it 3}& -2.206 &  2.635 &63.83 &63.89 & 0.41 &   0.463 &  0.470 & 2.6(22.62) & 0.13(0.0149) \\ 
\multicolumn{9}{l}{\bf G41.16$-$00.20} \\
{\it 1}&-47.636 & 19.203 &55.99 &55.98 & 0.28 &   0.298 &  0.309 & 0.8(6.96) & 0.12(0.0138) \\ 
{\it 2}&  0.000 &  0.000 &61.78 &61.80 & 0.24 &   0.647 &  0.621 & 0.8(6.96) & 0.42(0.0482) \\ 
\multicolumn{9}{l}{\bf G41.226$-$00.197}\\
{\it 1}& 20.176 & -2.239 &57.33 &57.50 & 0.72 &   0.958 &  0.513 & 4.9(42.63) & 0.09(0.0103) \\ 
&   &   &  &57.32 & 0.33 &   &  0.512 &   &  \\ 
{\it 2}& 23.727 & -2.827 &57.77 &57.75 & 0.16 &   0.891 &  0.463 & 3.7(32.19) & 0.07(0.0081) \\ 
&   &   &  &57.80 & 0.47 &   &  0.453 &   &  \\ 
{\it 3}&  0.000 &  0.000 &55.40 &55.47 & 0.78 &   1.977 &  1.030 & 2.1(18.27) & 0.54(0.0621) \\ 
&   &   &  &55.43 & 0.17 &  &  1.065 &   &  \\ 
{\it 4}& -2.196 & -2.924 &57.07 &57.01 & 0.68 &   1.559 &  1.548 & 3.1(26.97) & 0.42(0.0483) \\ 
{\it 5}& -1.131 & -4.055 &58.03 &58.06 & 0.19 &   0.396 &  0.430 & 0.7(6.09) & 0.41(0.0471) \\ 
{\it 6}&-14.881 &-34.934 &61.55 &61.54 & 0.32 &   0.370 &  0.371 & 0.7(6.09) & 0.48(0.0552) \\ 
{\it 7}&-24.354 &-30.431 &62.42 &62.42 & 0.38 &   1.079 &  0.551 & 0.5(4.35) & 1.14(0.1310) \\ 
&  &   &  &62.31 & 0.82 &   &  0.547 &  &  \\ 
{\it 8}&-23.849 &-30.070 &62.78 &62.73 & 0.32 &   1.268 &  1.295 & 0.7(6.09) & 0.52(0.0598) \\ 
\multicolumn{9}{l}{\bf G41.348$-$00.136}\\
{\it 1}& 41.993 &-27.277 & 7.09 & 7.36 & 0.39 &   2.140 &  0.922 & 4.6(53.36) & 0.17(0.0147) \\ 
&   &   &  & 7.10 & 0.24 &   &  1.879 &   &  \\ 
{\it 2}& 54.675 &-21.390 & 7.79 & 7.81 & 0.21 &   0.348 &  0.352 & 4.0(46.40) & 0.04(0.0035) \\ 
{\it 3}& 39.270 &-19.666 & 8.06 & 8.07 & 0.23 &   0.769 &  0.776 &  0.7(8.12) & 0.40(0.0345) \\ 
{\it 4}& 38.899 &-19.831 & 8.50 & 8.49 & 0.32 &   1.707 &  1.692 & 2.1(24.36) & 0.25(0.0216) \\ 
{\it 5}& 39.591 &-32.498 & 9.11 & 9.03 & 0.45 &   0.809 &  0.779 & 2.8(32.48) & 0.22(0.0190) \\ 
{\it 6}&-10.171 &  5.827 &11.66 &11.62 & 0.34 &   5.245 &  5.318 & 1.9(22.03) & 0.28(0.0241) \\ 
{\it 7}&  3.920 & -2.638 &11.92 &11.91 & 0.40 &   3.693 &  3.695 & 4.4(51.03) & 1.00(0.0862) \\ 
{\it 8}&  0.000 &  0.000 &12.27 &12.22 & 0.68 &   7.309 &  5.661 & 2.5(28.99) & 0.00(0) \\ 
&  &  & &12.33 & 0.23 &  &  2.027 &   &  \\ 
{\it 9}&  0.534 &  0.462 &12.97 &12.93 & 0.48 &   2.882 &  2.967 & 1.1(12.75) & 0.56(0.0483) \\ 

\multicolumn{9}{l}{\bf G43.165$+$00.013 (W49N)}\\
{\it 1}& 25.162 & 27.858 & 8.32 & 8.31 & 0.35 &   0.879 &  0.821 &  6.9(76.66)   & 0.06(0.0054) \\ 
{\it 2}& 42.184 & 29.218 & 8.32 & 8.32 & 0.45 &   1.200 &  1.196 &  5.4(59.99)   & 0.10(0.0090) \\ 
{\it 3}& 39.325 & 24.898 & 8.67 & 8.58 & 0.49 &   0.882 &  0.873 &  6.0(66.67)   & 0.07(0.0063) \\ 
{\it 4}&  0.000 &  0.000 & 9.29 & 9.27 & 0.28 &   6.190 &  6.157 &  5.4(59.99)   & 0.10(0.0090) \\ 
{\it 5}& 12.848 &  6.882 & 9.55 & 9.50 & 0.18 &   2.822 &  1.851 & 10.2(113.32)  & 0.08(0.0072) \\ 
 &  &   &  & 9.63 & 0.55 &   &  1.441 &  &  \\ 
{\it 6}&-51.298 &-127.361 &15.61 &15.60 & 0.22 &   0.305 &  0.306 & 0.7(77.77)  & 0.25(0.0225) \\ 
{\it 7}&-65.457 &-117.248 &17.36 &17.31 & 0.59 &   0.640 &  0.597 & 7.1(78.88)  & 0.06(0.0054) \\ 
{\it 8}& 18.732 &-142.429 &17.89 &17.90 & 0.46 &   0.861 &  0.777 & 4.5(50.00)  & 0.10(0.0090) \\ 
{\it 9}& 14.136 &-137.454 &18.42 &18.45 & 0.58 &   1.123 &  1.122 & 8.3(92.21)  & 0.06(0.0054) \\ 
{\it 10}&-24.526 &-124.971 & 18.77 &18.78 & 0.42 &   0.744 &  0.732& 0.6(6.67)  & 0.42(0.0038) \\ 
{\it 11}& 15.225 &-119.198 &19.03 &19.05 & 0.21 &   1.343 &  1.366 &1.5(16.67)  & 0.18(0.0162) \\ 
{\it 12}& 26.916 &-96.562 &19.56 &19.59 & 0.26 &   2.732 &  2.824 & 2.6(28.89)  & 0.17(0.0153) \\ 

\multicolumn{9}{l}{\bf G43.171$+$00.004 (W49N)}\\
{\it 1}& 36608.800 &3169.290 &19.03 &19.13 & 0.78 &   0.809 &  0.434 & 5.7(63.33) & 0.06(0.0054) \\ 
&   &   &  &19.01 & 0.15 &    &  0.423 &   &  \\ 
{\it 2}&36604.600 &3202.250 &19.03 &19.06 & 0.14 &   1.297 &  0.379 & 5.4(59.99) & 0.08(0.0072) \\ 
&   &   &  &19.04 & 0.37 &    &  0.955 &   &  \\ 
{\it 3}&36644.900 &3289.390 &20.26 &20.25 & 0.69 &   0.479 &  0.152 & 4.9(54.44) & 0.07(0.0063) \\ 
&   &   &  &20.25 & 0.69 &    &  0.152 &   &  \\ 
&   &   &  &20.28 & 0.18 &    &  0.167 &   &  \\ 
{\it 4}&36708.100 &3156.490 &21.75 &22.04 & 0.95 &   0.393 &  0.252 & 2.1(23.33) & 0.13(0.0117) \\ 
&   &   &  &22.08 & 1.01 &    &  0.263 &   &  \\ 
{\it 5}&36708.000 &3158.330 &21.93 &21.96 & 0.33 &   0.615 &  0.323 & 3.1(34.44) & 0.11(0.0099) \\ 
&   &   &  &21.97 & 0.30 &    &  0.323 &   &  \\ 
\multicolumn{9}{l}{\bf G43.167$-$00.004 (W49N)}\\
-- &  & & & & & & & & \\
\multicolumn{9}{l}{\bf G43.149$+$00.013 (W49N)}\\
{\it 1}&-27172.400 &-51712.000 &13.24 &13.25 & 0.23 &   1.449 &  1.480 & 3.0(33.33) & 0.09(0.0081)\\ 
{\it 2}&-27104.200 &-51740.700 &13.59 &13.56 & 0.19 &   0.785 &  0.815 & 1.5(16.67) & 0.12(0.0108) \\ 
{\it 3}&-27026.700 &-51822.400 &14.03 &14.04 & 0.24 &   0.416 &  0.423 & 3.9(43.33) & 0.05(0.0045) \\ 
\\
\\
\multicolumn{9}{l}{\bf G45.467$+$00.053}\\
{\it 1}&  4.543 & -6.721 &56.54 &56.18 & 0.66 &   0.570 &  1.336 & 3.3(23.76) & 0.08(0.0111) \\ 
{\it 2}&  0.000 &  0.000 &56.01 &56.19 & 0.64 &   3.024 &  1.768 & 3.8(27.36) & 0.23(0.0319) \\ 
&   &   &  &56.01 & 0.28 &    &  1.560 &   &  \\ 
{\it 3}&  2.652 & -3.864 &56.71 &56.71 & 0.22 &   0.685 &  0.359 & 1.6(11.52) & 0.22(0.0306) \\ 
&  &  &   &56.72 & 0.84 &   &  0.322 &   &  \\ 
{\it 4}&  0.555 & -5.502 &57.50 &57.45 & 0.51 &   0.972 &  0.969 & 3.1(22.32) & 0.20(0.0278) \\ 
{\it 5}&  7.982 & -5.674 &57.50 &57.49 & 0.23 &   0.613 &  0.625 & 2.0(14.40) & 0.09(0.0125) \\ 
{\it 6}& 14.815 & -5.892 &57.86 &57.84 & 0.24 &   0.558 &  0.608 & 1.3(9.36) & 0.21(0.0292) \\ 
{\it 7}&  7.530 &-14.983 &58.12 &58.14 & 0.33 &   1.877 &  1.900 & 0.9(6.48) & 0.69(0.0958) \\ 
{\it 8}&  5.142 &-12.853 &59.17 &59.11 & 0.33 &   0.348 &  0.337 & 3.3(23.76) & 0.13(0.0181) \\ 
\multicolumn{9}{l}{\bf G45.473$+$00.134}\\
{\it 1}&-718.044 &-371.031 &62.25 &62.30 & 0.25 &   1.720 &  1.808 & 1.7(11.73) & 0.20(0.0290) \\ 
{\it 2}&-38.950 &403.500 &59.62 &59.63 & 0.20 &   0.679 &  0.681 & 3.0(20.70) & 0.09(0.0130) \\ 
{\it 3}&-35.561 &-23.101 &64.27 &64.29 & 0.32 &   0.632 &  0.631 & 1.4(9.66) & 0.25(0.0362) \\ 
{\it 4}&-35.660 & -9.649 &65.06 &65.05 & 0.37 &   0.237 &  0.252 & 4.9(33.81) & 0.07(0.0101) \\ 
{\it 5}&  2.491 &-14.945 &66.29 &66.30 & 0.22 &   1.427 &  1.467 & 0.8(5.52)) & 0.45(0.0652) \\ 
{\it 6}&  0.000 &  0.000 &65.76 &65.90 & 0.51 &   3.669 &  1.553 & 1.1(7.59) & 0.96(0.1391) \\ 
&  &  &  &65.75 & 0.35 &   &  2.501 &   &  \\ 
\multicolumn{9}{l}{\bf G45.493$+$00.126}\\
{\it 1}&  0.000 &  0.000 &57.23 &57.47 & 0.52 &   4.357 &  1.270 & 1.8(12.78) & 0.34(0.0479) \\ 
&   &   &  &57.25 & 0.29 &    &  3.714 &   &  \\ 
{\it 2}&  2.822 & -0.728 &57.94 &57.91 & 0.39 &   1.335 &  1.353 & 0.8(5.68) & 0(0) \\ 
\multicolumn{9}{l}{\bf G59.782$+$00.065}\\
{\it 1}&-720.339 &-296.368 &14.56 &14.57 & 0.33 &   3.617 &  3.689 & 0.4(0.86) & 1.28(0.5926) \\ 
{\it 2}&-721.714 &-325.084 &15.62 &15.58 & 0.62 &   0.510 &  0.508 & 1.2(2.59) & 0.37(0.1713) \\ 
{\it 3}&-733.255 &-306.394 &15.53 &15.56 & 0.30 &   1.810 &  1.852 & 2.4(5.184) & 0.15(0.0694) \\ 
{\it 4}&-739.724 &-249.439 &15.70 &15.64 & 0.50 &   0.743 &  0.734 & 1.1(2.376) & 0.32(0.1482) \\ 
{\it 5}&-716.508 &-300.663 &15.53 &15.57 & 0.56 &   6.649 &  6.413 & 2.2(4.75) & 0.51(0.2361) \\ 
{\it 6}&-693.358 &-352.404 &16.85 &16.86 & 0.32 &   4.327 &  4.319 & 1.1(2.376) & 0.32(0.1482) \\ 
{\it 7}&-695.021 &-353.759 &17.20 &17.20 & 0.34 &   2.237 &  2.378 & 0.6(1.30) & 0.63(0.2917) \\ 
{\it 8}&-709.908 &-345.120 &19.83 &19.80 & 0.20 &   4.774 &  5.197 & 0.6(1.30) & 0.48(0.2222) \\ 
{\it 9}&-830.068 &-144.758 &20.01 &19.96 & 0.28 &  10.889 & 11.636 & 1.5(3.24) & 0.35(0.1620) \\ 
{\it 10}&-700.456 &-357.063 &20.09 &20.09 & 0.35 &   5.700 &  5.812 & 1.7(3.67) & 0.26(0.1204) \\ 
{\it 11}&-810.734 &-163.106 &19.74 &19.75 & 0.22 &  14.435 &  9.077 & 7.8(16.85) & 0.09(0.0417) \\ 
&  &   &  &19.64 & 0.42 & &  6.657 &  &  \\ 
{\it 12}&-816.398 &-159.881 &20.18 &20.18 & 0.57 &   1.248 &  1.193 & 6.1(13.18) & 0.07(0.0324) \\ 
{\it 13}&-698.311 &-371.858 &20.71 &20.74 & 0.55 &   1.241 &  1.205 & 5.0(10.80) & 0.14(0.0648) \\ 
{\it 14}&-697.668 &-378.427 &21.50 &21.43 & 0.54 &   1.602 &  1.633 & 3.4(7.34) & 0.13(0.0602) \\ 
{\it 15}&147.224 & 52.502 &15.70 &15.61 & 0.53 &   3.913 &  3.964 & 1.8(3.89) & 0(0) \\ 
{\it 16}&289.167 &-90.327 &17.20 &17.20 & 0.28 &   2.688 &  2.544 & 6.6(14.26) & 0.07(0.0324) \\ 
{\it 17}&314.263 &-64.063 &18.78 &18.78 & 0.33 &   0.315 &  0.315 & 0.9(1.94) & 0.20(0.0926) \\ 
{\it 18}&349.850 &-55.566 &19.04 &19.05 & 0.28 &   1.298 &  1.310 & 0.9(1.94) & 0.29(0.1343) \\ 
{\it 19}&310.997 &-81.716 &19.13 &19.17 & 0.31 &   8.178 &  8.637 & 1.5(3.24) & 0.41(0.1898) \\ 
{\it 20}& 52.423 &  2.233 &23.43 &23.42 & 0.30 &   0.626 &  0.622 & 1.4(3.02) & 0(0) \\ 
{\it 21}& 52.565 & -1.239 &24.13 &24.11 & 0.31 &   1.446 &  1.457 & 1.9(4.10) & 0.24(0.1111) \\ 
{\it 22}& 51.951 & -7.046 &24.75 &24.69 & 0.25 &  14.610 & 15.646 & 0.3(0.65) & 0(0) \\ 
{\it 23}&  0.000 &  0.000 &27.03 &27.03 & 0.36 &  37.071 & 37.357 & 4.9(10.58) & 0.16(0.0741) \\ 
{\it 24}& 18.630 &  4.624 &27.38 &27.33 & 0.28 &   6.838 &  7.317 & 2.1(4.54) & 0.26(0.1204) \\ 
\end{longtable}
\tablefoot{$^a$ The values of the projected lengths and the velocity gradients in brackets 
are calculated for the exact distances as described in Sect.~4.1 and listed
in Table~\ref{source-prop}.}

\begin{table*}
 \centering
  \caption{Auxiliary inputs to the SED models. Data are taken from catalogs as explained in the text
(Sect.~4.2). The upper and lower limits are marked by$\downarrow$ and $\uparrow$, respectively.}
  \begin{tabular}{lllcccc}
 \hline\hline
 &\multicolumn{2}{c}{Band}  & G37.753$-$00.189 & G40.282$-$00.219 & G40.425$+$00.700 & G41.123-00.220\\
 & & ($\mu$m)  & (Jy) & (Jy) & (Jy) & (Jy)\\
\hline
UKIDSS & J& 1.248 & 0.00007& 0.00017& 0.00094& ...\\
       & H& 1.631 & 0.0002 & 0.0029 & 0.0023 & ...\\
       & K& 2.201 & 0.0003 & 0.015  & 0.0027 & 0.00025\\
Spitzer & IRAC [1] & 3.6 & 0.00213 & 0.129 & 0.010 & 0.0012\\
        & IRAC [2] & 4.5 & 0.01048 & 0.252 & 0.025 & 0.004\\
        & IRAC [3] & 5.8 & 0.02511 & 0.2248& 0.101 & 0.008\\
        & IRAC [4] & 8.0 & ... & 0.2885 & 0.174 & ...\\
MSX   & A & 8.28& ...& 0.0634& 1.330 & 0.262$\downarrow$\\
      & C &12.13& ...& ...   & 2.248 &...\\
      & D &14.64& ...& 1.031 & 2.823 &...\\
      & E &21.34& ...& 7.153 & 9.300 &...\\
MIPS & [1] & 24 & 0.841 & 5.9$\uparrow$& 6$\uparrow$ & 0.15\\
     & [2] & 70 &  ...  & ... & 322.94 &...\\
SCUBA & & 450 & 1.32 & ...& ...&...\\
      & & 850 & 0.51 & ...& ...&...\\
LABOCA& & 870 & ... & 12.52 & ...&...\\
BOLOCAM& & 1100 & 2.122 & 5.537 &...& 3.039\\
IRAM  & & 1200 & ...& 3.156&...&...\\
Distance range & (kpc) & &7.9--9.7 & 4.3--5.8 & 10.3--12.5 & 7.8--9.6 \\
\\
\hline
\hline
 &\multicolumn{2}{c}{Band} & G41.16$-$00.20 & G41.226$-$00.197 &G41.348-00.136 &\\
 & & ($\mu$m)  & (Jy) & (Jy) & (Jy) & \\
\hline
UKIDSS & H& 1.631 & 0.00012 & 0.00013&  ...& \\
       & K& 2.201 & 0.0002  & 0.0011 & 0.000056& \\
Spitzer & IRAC [1] & 3.6 &0.0003 & 0.0044& ...& \\
        & IRAC [2] & 4.5 &0.0013 & 0.0081& 0.0056& \\
        & IRAC [3] & 5.8 &0.0012 &   ... & 0.0119 & \\
        & IRAC [4] & 8.0 &0.0047 &   ... & 0.0205 &\\
MSX   & A & 8.28& ...& 0.689$\downarrow$ & ...&\\
      & C &12.13& ...& 1.063$\downarrow$ & ...&\\
      & D &14.64& ...& 0.971$\downarrow$ & ...&\\
MIPS & [1] & 24 & 0.1877 & 0.260 & ...&\\
SCUBA & & 450 & ...&...& 428.57 &\\
      & & 850 & ...&...& 1.5 &\\
LABOCA& & 870 & ... &1.59 & ...& \\
BOLOCAM& & 1100 & ...& 1.268 & 0.407&\\
IRAM  & & 1200 & ...& 0.264 & \\
Distance range & (kpc) & & 7.8--9.6& 7.8--9.6 & 10.4--12.8& \\
\\
\hline
\hline
 &\multicolumn{2}{c}{Band}  & G45.467$+$00.053 &G45.473$+$00.134 &G45.493$+$00.126 & G59.782$+$00.065 \\
 & & ($\mu$m)  & (Jy) & (Jy) & (Jy) & (Jy)\\
\hline
UKIDSS & J& 1.248 & ... & 0.00015 & ...&  0.00019\\
       & H& 1.631 & 0.00018& 0.00042 & 0.00003 & 0.0019\\
       & K& 2.201 &  0.00091& 0.002 & ...& 0.0319 \\
Spitzer & IRAC [1] & 3.6 & 0.0056& ... & 0.00093  & 2.092\\
        & IRAC [2] & 4.5 & 0.0328& 0.062 & 0.0046 & 7.373\\
        & IRAC [3] & 5.8 & 0.0880 & 0.132 & 0.0035 & 4.436\\
        & IRAC [4] & 8.0 & 0.0874 & ... & ... & ...\\
MSX   & A & 8.28& ...& 12.43 & ... & 5.325\\
      & C &12.13& ...& 32.030& ... & 8.528\\
      & D &14.64& 0.605 & 36.41& ... & 14.911\\
      & E &21.34& 1.692 & 172.70& ... & 47.399\\
MIPS & [1] & 24 &  10.0$\uparrow^3$ &...& ... &\\
Herschel & & 70 & ...& ...& ...& 206.04\\
         & & 250& ...& ...& ...& 459.70\\
         & & 350& ...& ...& ...& 232.26\\
         & & 500& ...& ...& ...& 111.45\\
SCUBA & & 450 & ...&...& ... & 19.93\\
      & & 850 & 40.12& 15.19 & 12.07 & 16.92\\
BOLOCAM& & 1100 & 6.9 & 9.457 & ... & 10.516\\
SIMBA  & & 1200 & ...& ...& ... & 4.7\\
Distance range & (kpc) & & 5.9--7.9  & 5.9--7.7 & 5.9--7.8 & 2.0--2.4\\
\hline
\label{sedin}
\end{tabular}
\end{table*}
\end{appendix}

\end{document}